\documentclass{article}
\usepackage[german, english]{babel}
\usepackage[a4paper, left=2.5cm, right=2.5cm, top=3.5cm, bottom=3cm]{geometry}
\usepackage{amssymb}
\usepackage{amsmath}
\usepackage{bbm}
\usepackage{amscd}
\usepackage{latexsym}
\usepackage{graphicx}
\usepackage{ifthen}
\usepackage{epsfig}
\usepackage{setspace}
\usepackage{subcaption}
\usepackage{dsfont}
\usepackage{mathrsfs}
\usepackage{mathabx}
\usepackage{color}
\usepackage{xcolor}

\DeclareMathAlphabet{\mathpzc}{OT1}{pzc}{m}{it}

\catcode`@=11
\renewcommand{\section}{\@startsection{section}{1}{0pt}{\medskipamount}
{\medskipamount}{\large\bf}}
\numberwithin{equation}{section}
\catcode`@=12

\newcommand{\Z}{\mathds Z}
\newcommand{\C}{\mathds C}

\newcommand{\R}{\mathds R}

\newcommand{\al}{\alpha}
\newcommand{\zb}{{\bar{z}}}
\newcommand{\wb}{{\bar{w}}}
\newcommand{\pb}{{\bar{p}}}
\newcommand{\qb}{{\bar{q}}}
\newcommand{\rb}{{\bar{r}}}
\newcommand{\tF}{\widetilde{F}}
\newcommand{\ve}{\varepsilon}
\newcommand{\s}{\sigma}
\newcommand{\z}{\zeta}

\def\im{\mathrm{i}}
\def\ep{\mathrm{e}}
\def\pa{\partial}
\def\diff{\mathrm{d}}
\def\tr{\mathrm{tr}}
\def\sfrac#1#2{{\textstyle\frac{#1}{#2}}}
\def\]{\right]}
\def\[{\left[}
\def\){\right)}
\def\({\left(}
\def\>{\rangle}
\def\<{\langle}
\def\+{\dagger}

\def\={\ =\ }
\def\+{\,+\,}
\def\ph{\phantom{-}}
\def\rs{\!\!}
\newcommand{\unity}{\mathbbm{1}}
\newcommand{\hateq}{\mathrel{\widehat{=}}}
\def\und{\quad\textrm{and}\quad}
\def\with{\quad\textrm{with}\quad}
\def\for{\quad\textrm{for}\quad}

\begin{document}

\title{\bf\huge A hyperbolic Kac--Moody Calogero model}
\date{~}

\author{
{\Large Olaf Lechtenfeld$^{\,\star}$} 
\\[12pt]
{\Large with a joint appendix with Don Zagier$^{\,\diamond}$}
\\[24pt]
{\Large ${\,}^\star$}
Institut f\"ur Theoretische Physik and\\  \quad
Riemann Center for Geometry and Physics\\ \quad
Leibniz Universit\"at Hannover \\ \quad
Appelstra{\ss}e 2, 30167 Hannover, Germany
\\[12pt]
{\Large ${\,}^\diamond$}
Max-Planck-Institut f\"ur Mathematik\\ \quad
Vivatsgasse 7, 53111 Bonn, Germany
\\[24pt]
} 

\clearpage
\maketitle
\thispagestyle{empty}

\begin{abstract}
\noindent\large
A new kind of quantum Calogero model is proposed, based on a hyperbolic Kac--Moody algebra.
We formulate nonrelativistic quantum mechanics on the Minkowskian root space 
of the simplest rank-3 hyperbolic Lie algebra $AE_3$ with an inverse-square potential
given by its real roots and reduce it to the unit future hyperboloid.
By stereographic projection this defines a quantum mechanics on the Poincar\'e disk
with a unique potential. Since the Weyl group of $AE_3$ is a $\Z_2$~extension of the
modular group PSL(2,$\Z$), the model is naturally formulated on the complex upper half plane,
and its potential is a real modular function. We present and illustrate the relevant features 
of $AE_3$, give some approximations to the potential and rewrite it as an (almost everywhere convergent) 
Poincar\'e series. The standard Dunkl operators are constructed and investigated
on Minkowski space and on the hyperboloid. In the former case
we find that their commutativity is obstructed by rank-2 subgroups of hyperbolic type 
(the simplest one given by the Fibonacci sequence), casting doubt on the integrability of the model.
An appendix with Don Zagier investigates the computability of the potential.
We foresee applications to cosmological billards and to quantum chaos.
\end{abstract}

\newpage
\setcounter{page}{1} 

\section{Introduction}
\noindent
To every finite Coxeter group of rank~$n$ one can associate 
a (classical and quantum) maximally superintegrable mechanical system 
known as a (rational) Calogero--Moser model (Calogero model for short)
living in $2n$-dimensional phase space with momenta~$p_i$ and
coordinates~$x^i$ collected into~$x\in\R^n$~\cite{Cal,OlPe1,OlPe2}.
It is determined by the Hamiltonian
\begin{equation}
H \= \tfrac12\sum_{i=1}^n p_i^2\ +\ V(x)
\qquad\with\quad
V(x) \= \tfrac12\sum_{\al\in{\cal R}} g_\al(g_\al{-}\hbar)\,\frac{\al{\cdot}\al}{2(\al{\cdot}x)^2}\ ,
\end{equation}
where `${\cdot}$' denotes the standard Euclidean scalar product in~$\R^n$,
and the sum runs over the root system~$\cal R$ consisting of all nonzero roots~$\al$
belonging to the Coxeter-group reflections~\footnote{
We sum over both positive and negative roots and correct the overcount
due to the pair $(\al,-\al)$ by a factor of~$1/2$.}
\begin{equation}
s_\al:\ \R^n\to\R^n \qquad\textrm{via}\qquad s_\al x \= x - 2\tfrac{x{\cdot}\al}{\al{\cdot}\al}\al\ .
\end{equation}
The real coupling constants $g_\al$ are constant on each Weyl-group orbit,
so in an irreducible simply-laced case they all agree, $g_\al{=}g$.
For the classical Hamiltonian, $\hbar{=}0$, while in the quantum case we represent
the momenta by differential operators,
\begin{equation}
p_i \= \tfrac{\hbar}{\im}\tfrac{\pa}{\pa x^i} \ =:\  -\im\hbar\,\pa_i 
\qquad\textrm{such that}\qquad [x^i,p_j] = \im\hbar\,\delta^i_{\ j}\ .
\end{equation}
Henceforth we set $\hbar=1$ for convenience.

There exist a variety of generalizations for these models,
but we like to mention only their restriction to the unit sphere~$S^{n-1}$
given by $x{\cdot}x=1$,
\begin{equation} \label{hypHam}
H_\Omega \= -\tfrac12 L^2 + U(\vartheta) \quad\with\quad
L^2=\sum_{i<j}(x^i\pa_j{-}x^j\pa_i)^2 \quad\und\quad U(\vartheta)=x{\cdot}x\, V(x)\ ,
\end{equation}
which has been named the (spherical) angular Calogero model
\cite{Feigin,HNY,FLP}.
In the quantum version, $L^2$ is the (scalar) Laplacian on~$S^{n-1}$,
and the potential~$U$ depends only on its angular coordinates
$\vartheta=\{\vartheta_1,\ldots,\vartheta_{n-1}\}$.
Since $V(x)$ is singular at the mirror hyperplanes $\al{\cdot}x{=}0$ of the Coxeter group,
$U(\vartheta)$ blows up at their intersection with the unit sphere.
The full as well as the angular Calogero model and their generalization have a rich history
as paradigmatic many-body integrable models~(for a review, see~\cite{Polyrev,scholarpedia}).

Generalizing to infinite Coxeter groups of the affine type renders the coordinates periodic,
$x\in T^n$, which turns rational Calogero models into Sutherland models~\cite{Suth}.
However, to the author's knowledge, hyperbolic Coxeter groups~\cite{Davis} have not been employed
to this purpose, except for~\cite{fring1} on Toda field theories and~\cite{fring2}\footnote{
which appeared some time after the submission of this article to the arXiv}
on a more general class of Calogero models. 

In the present paper, we propose a rational Calogero model based on one of
the simplest hyperbolic Coxeter groups, namely the paracompact right triangular hyperbolic group
labelled by $[p,q,r]=[2,3,\infty]$ and Coxeter--Dynkin diagram \
$\bullet\underset{\infty}{\textrm{\textemdash}}\!\bullet\!\underset{3}{\textrm{\textemdash}}\bullet$
(see Figure~1).
This happens to be the Weyl group of the simplest hyperbolic rank-3 Lie algebra~$AE_3$, 
a double extension of~$A_1\equiv sl_2$~\cite{DHN}.\footnote{
Other names for this algebra are $\mathpzc{F}$~\cite{FF}, $H_3$~\cite{Kac}, 
$HA_1^{(1)}$~\cite{Kang1,Kang2} or $A_1^{{\wedge}\!{\wedge}}$~\cite{DHJN}.}
Its root space is of Lorentzian signature, which we take as~$(-,+,+)$.
\begin{figure}[h!tbp]
\centering
\vspace{-1.8cm}
\includegraphics[width = 0.6\paperwidth]{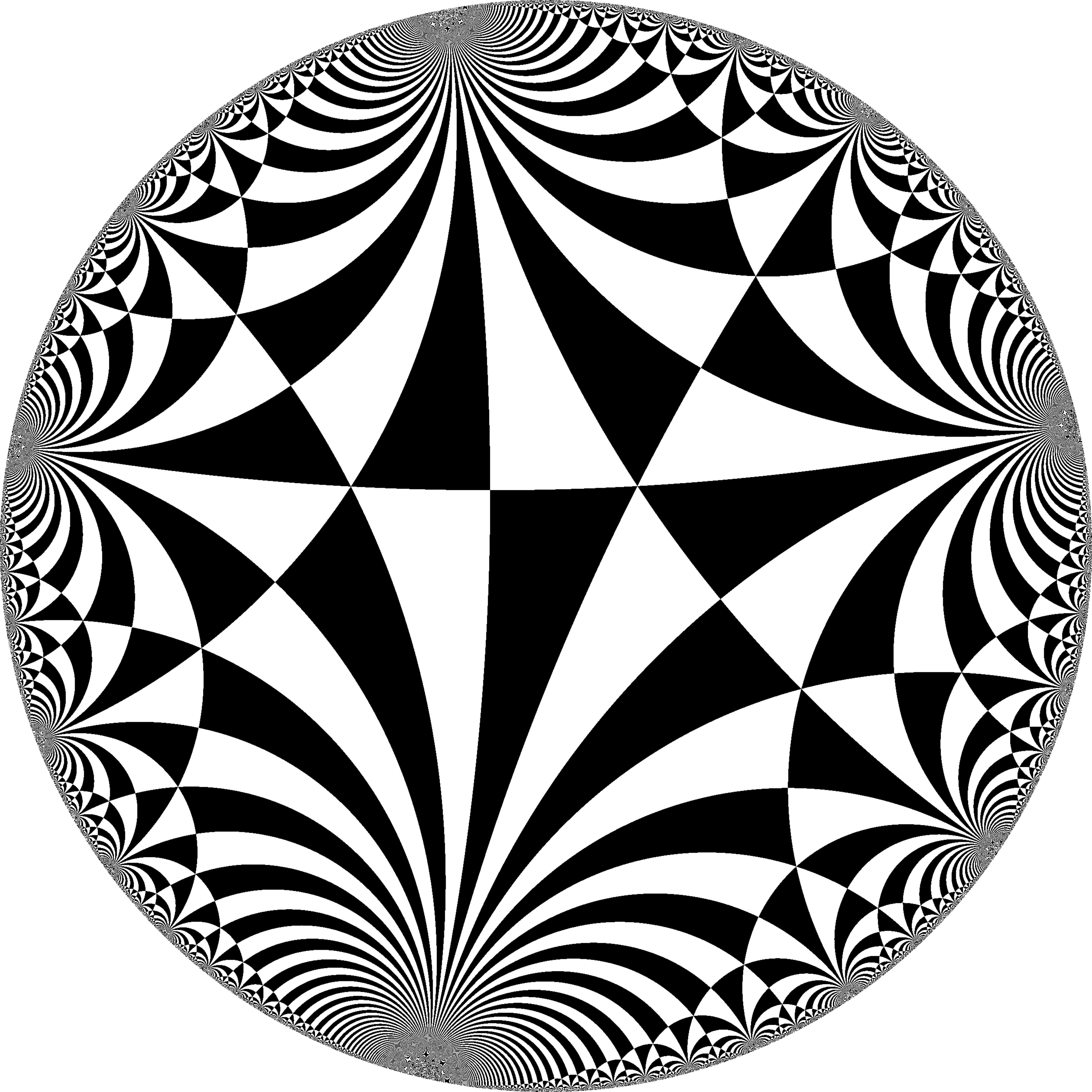} 
\caption{Poincar\'e disk model of fundamental domain triangles for 
the hyperbolic Coxeter group $[2,3,\infty]$}
\end{figure}
Denoting the phase-space coordinates by $(x^\mu,p_\mu)$ for $\mu=0,1,2$,
the Minkowski metric by~$\eta_{\mu\nu}$ and the Lorentzian scalar product again by `$\cdot$', 
the Hamiltonian then has the form 
(employing the Einstein summation convention and pulling the coupling out of the potential)
\begin{equation}
H \= \tfrac12 \eta^{\mu\nu} p_\mu p_\nu +\ g(g{-}1)\,V(x)
\= \tfrac12\bigl(-p_0^2+p_1^2+p_2^2\bigr)\ +\ 
\tfrac12\sum_{\al\in{\cal R}}\frac{g(g{-}1)}{(\al\cdot x)^2}\ ,
\end{equation}
where the sum is restricted to the set~$\cal R$ of {\it real\/} roots, which we normalize to 
$\al{\cdot}\al=2$.
Since $\cal R$ for $AE_3$ decomposes into {\it two\/} Weyl orbits 
${\cal R}_+$ and ${\cal R}_-$~\cite{CCFP},
we may actually split the potential into two pieces and weigh them individually. 
However, for the sake of simplicity we keep the couplings equal for this paper, $g_+{=}g_-{=}g$.
We also do not consider the possible inclusion of imaginary roots in the potential.

Like the Euclidean theory can be reduced to the unit sphere, the Minkowskian variant
can be restricted to the one-sheeted hyperboloid ($x{\cdot}x=1$) 
or (one sheet of) the two-sheeted hyperboloid ($x{\cdot}x=-1$). 
In order to produce a model on a Riemannian manifold (with Euclidean signature), 
we consider the future hyperboloid~$H^2$ given by $x{\cdot}x=-1$ and $x^0>1$.
Let us parametrize the Minkowski future by
\begin{equation}
x^0 = r\,\cosh\theta\ ,\quad x^1 = r\,\sinh\theta\,\cos\phi\ ,\quad x^2 = r\,\sinh\theta\,\sin\phi
\quad\with r\in\R_{>0}\ ,\quad \theta\in\R_{\ge0}\ ,\quad\phi\in[0,2\pi]
\end{equation}
so that we may restrict to $r{=}1$ and obtain the quantum Hamiltonian
of a ``hyperbolic Calogero model''~\footnote{
not to be confused with the hyperbolic Calogero--Sutherland model~\cite{OlPe1,OlPe2}.}
\begin{equation} \label{HOm}
H_\Omega \=  -\tfrac12 L^2 + g(g{-}1)\,U(\theta,\phi) \quad\with\quad
L^2 = \pa_\theta^2+\coth\theta\,\pa_\theta+\textrm{csch}^2\!\theta\,\pa_\phi^2 
\quad\und\quad U(\theta,\phi)=r^2\,V(x)\ ,
\end{equation}
where $L^2$ is just the (scalar) Laplacian on~$H^2$.
Our task will be to compute and characterize the potential~$V$ respective~$U$.
\section{The real roots of the Kac--Moody algebra $AE_3$}
\noindent
In order to formulate the Calogero potential for the real roots of $AE_3$ 
we need to collect some facts about this simplest of hyperbolic Kac--Moody algebras
\cite{Kac,FF,DHN}.
Starting from its Cartan matrix,
\begin{equation}
A \= \begin{pmatrix} \ph 2 & -2 & \ph 0 \ph \\ -2 & \ph 2 & -1 \ph \\ \ph 0 & -1 & \ph 2 \ph \end{pmatrix}\ ,
\end{equation}
we parametrize the three simple roots $\al_\mu$ of length-square~2
in three-dimensional Minkowski space $\R^{1,2}$ with a Minkowski-orthonormal basis 
$\{e_\mu\}=\{e_0,e_1,e_2\}$,
\begin{equation}
e_\mu{\cdot}\,e_\nu \= \eta_{\mu\nu} \quad\for\quad (\eta_{\mu\nu})=\textrm{diag}(-1,+1,+1)\ ,
\end{equation}
via
\begin{equation}
\al_0\=\sqrt{2}\,\bigl(\tfrac{1}{\sqrt{3}}e_0-e_1-\tfrac{1}{\sqrt{3}}e_2\bigr)\ ,\qquad
\al_1\=\sqrt{2}\,e_1\ ,\qquad
\al_2\=\sqrt{2}\,\bigl(-\tfrac12 e_1 +\tfrac12\sqrt{3}\,e_2\bigr)\ .
\end{equation}
For symmetry reasons we add the non-simple root
\begin{equation}
\al_3 \= -\al_1{-}\al_2 \= \sqrt{2}\bigl(-\tfrac12 e_1 -\tfrac12\sqrt{3}\,e_2\bigr)\ ,
\end{equation}
so that the overextended simple root can be rewritten as
\begin{equation} \label{al0}
\al_0 \= \sqrt{\tfrac23}\,e_0 -\tfrac23\,\al_1 +\tfrac23\,\al_3\ .
\end{equation}
The three roots $\al_i$, $i=1,2,3$, belong to an~$A_2$ subalgebra and obey the relations
\begin{equation}
\al_1+\al_2+\al_3=0\ ,\qquad
\al_i{\cdot}\,\al_i=2\ ,\qquad
\al_i{\cdot}\,\al_j=-1\ (i{\neq}j)\ ,\qquad
(\al_i-\al_j){\cdot}\,\al_k=0\ (i{\neq}k{\neq}j)\ .
\end{equation}

The real roots of $AE_3$ lie on the one-sheeted hyperboloid $x{\cdot}x{=}2$ 
and are given by
\begin{equation}
\al \= \ell\,\al_0 + m\,\al_1 + n\,\al_2 \quad\with\quad
\ell,m,n\in\Z \quad\und\quad \al\cdot\al=2\ ,
\end{equation}
where the length condition translates to the diophantine equation
\begin{equation} \label{dio}
(\ell-m)^2+n\,(n-m) \= 1\ .
\end{equation}
Since the roots come in pairs $(\al,-\al)$, it suffices to analyze $\ell\ge0$ only.
At any given ``level''~$\ell$ the solutions furnish (generically several) 
highest weights of the ``horizontal'' $A_2\equiv sl_3$ subalgebra 
plus their images under its $S_3$ Weyl-group action,
\begin{equation} \label{S3}
(m,n)\ \to\ (m,m{-}n)\ \to\ (2\ell{-}n,m{-}n)\ \to\ 
(2\ell{-}n,2\ell{-}m)\ \to\ (2\ell{-}m{+}n,2\ell{-}m)\ \to\ (2\ell{-}m{+}n,n)\ .
\end{equation}
Such a sextet of weights belongs to an $A_2$ representation with Dynkin labels
\begin{equation}
[\pb,\qb] \= [-2\ell{+}2m{-}n,{-}m{+}2n]  \qquad\Leftrightarrow\qquad
(m,n)_{\textrm{hw}} \= \tfrac13\,(4\ell{+}2\pb{+}\qb,2\ell{+}\pb{+}2\qb)
\end{equation}
for the highest-weight values $(m,n)_{\textrm{hw}}$ in~(\ref{S3}).
Figure~2 shows the distribution of the real roots for low levels.
\begin{figure}[h!tbp]
\centering
\includegraphics[width = 0.35\paperwidth]{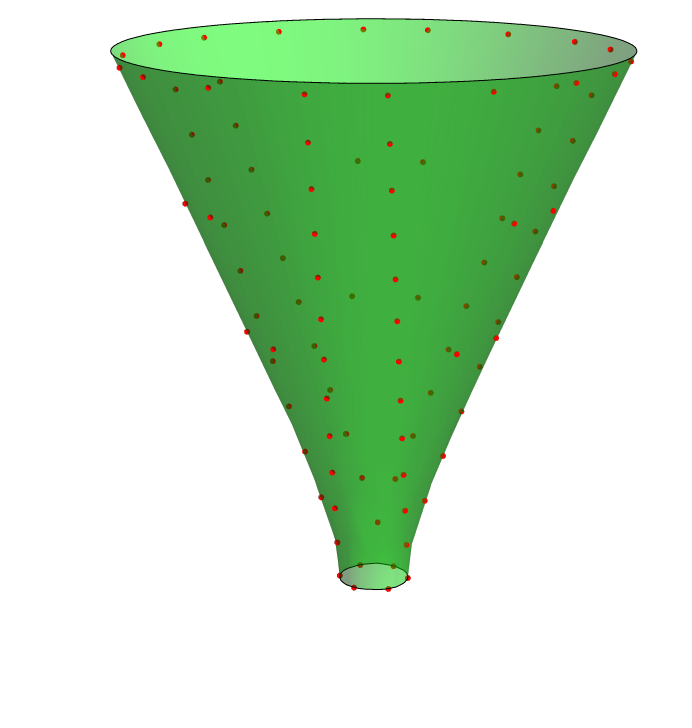} \qquad
\includegraphics[width = 0.35\paperwidth]{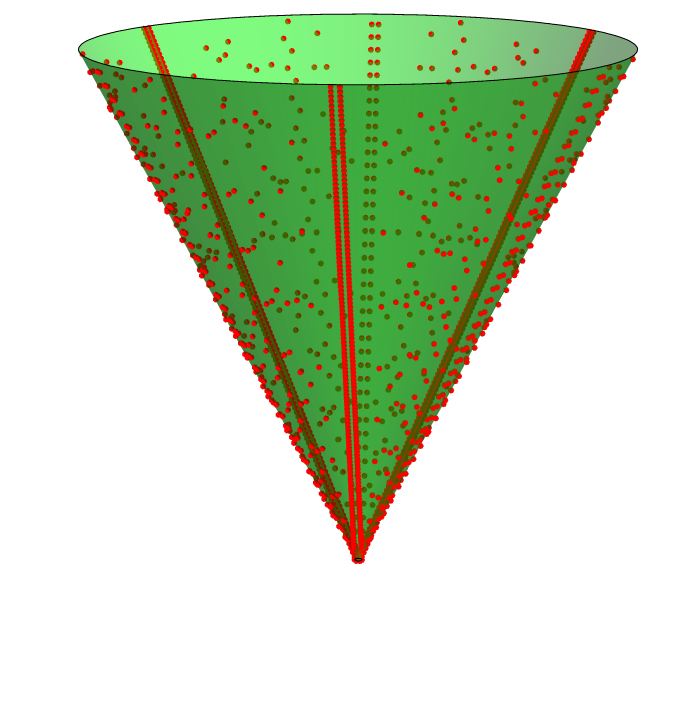} 
\vspace{-1cm}
\caption{Hyperboloid $x{\cdot}x=2$ (green) and real roots $\alpha$ (red)
for $0\le\ell\le12$ (left) and $0\le\ell\le122$ (right)}
\end{figure}

We obtain a parametrization more symmetric under spatial rotations
by replacing $\al_0$ with $e_0$ using~(\ref{al0}) and employing a
kind of barycentric coordinates,
\begin{equation}
x \= x^0\,e_0 + \bar{x}^i\,\al_i \qquad\with\quad 
\bar{x}^1{+}\bar{x}^2{+}\bar{x}^3=0\ .
\end{equation}
In these coordinates, the real roots take the form
\begin{equation}
\al \= \sqrt{\tfrac23}\,\ell\,e_0 + \bar{\al} \quad\with\quad
\bar{\al} \= \tfrac13\pb\,\al_1+\tfrac13\qb\,\al_2+\tfrac13\rb\,\al_3
\quad\und\quad \pb+\qb+\rb=0\ ,
\end{equation}
where $\pb$ and $\qb$ coincide with the Dynkin labels in case $\bar{\alpha}$ 
is a highest $A_2$~weight.
The $S_3$ Weyl group action simply permutes the coefficients $(\pb,\qb,\rb)$
and multiplies them with the sign of the permutation.
On a given level~$\ell$ the $A_2$ weights~$\bar{\al}$ all have the same length-square
$\bar{\al}{\cdot}\bar{\al}=2+\tfrac23\ell^2$.
On may translate the diophantine equation~(\ref{dio}) to the Dynkin labels and obtain
\begin{equation}
-\pb\qb-\pb\rb-\rb\pb \ \equiv\ \pb^2+\qb^2+\pb\,\qb \= \ell^2+3 
\qquad\textrm{as well as}\qquad 3 \mid \ell-\pb+\qb\ .
\end{equation}
The number of $A_2$ representations grows erratically with the level,
as displayed in Figure~3.
Two representations appear first at $\ell{=}6$, three at $\ell{=}12$, four at $\ell{=}30$,
eight at $\ell{=}72$ and so on.
\begin{figure}[h!tbp]
\centering
\includegraphics[width = 0.6\paperwidth]{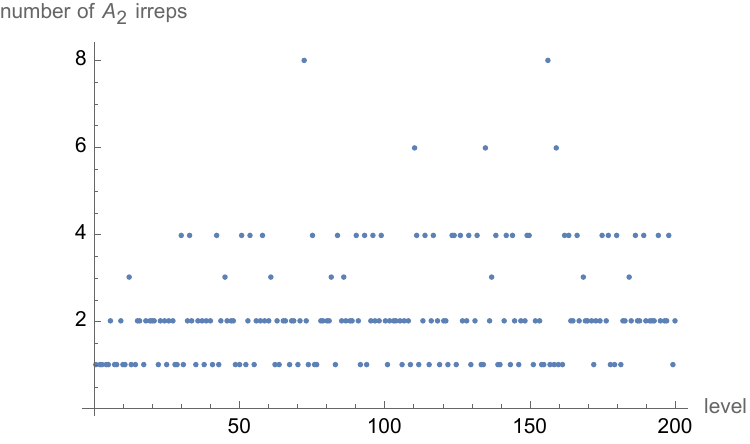} 
\caption{Multiplicity of $A_2$ representations occurring at level~$\ell$}
\end{figure}

At level zero one simply finds the adjoint representation,
\begin{equation}
\ell=0:\qquad (m,n)_{\textrm{hw}}=(1,1) 
\qquad\Leftrightarrow\qquad [\pb,\qb]=[1,1]\ .
\end{equation}
Although the number of solutions grows quickly with $\ell$ it is easy
to give a few infinite families of real roots (see also~\cite{DHN}),
\begin{equation} \label{series}
\begin{aligned}
\ell\ge1:\qquad &(m,n)_{\textrm{hw}}=(2\ell,\ell{+}1) 
\qquad&\Leftrightarrow&\qquad [\pb,\qb]=[\ell{-}1,2]\ ,\\
\ell=3k\ge3:\qquad &(m,n)_{\textrm{hw}}=(5k{+}1,4k) 
\qquad&\Leftrightarrow&\qquad [\pb,\qb]=[2,3k{-}1]\ ,\\
\ell=k(k{+}1)\ge6:\qquad &(m,n)_{\textrm{hw}}=(2k(k{+}1){-}1,k(k{+}2)) 
\qquad&\Leftrightarrow&\qquad [\pb,\qb]=[k^2{-}2,2k{+}1]\ ,
\end{aligned}
\end{equation}
where at $\ell{=}6$ the second and third series coincide for $k{=}2$, 
the $k{=}1$ solution is contained in the first series at $\ell{=}3$,
and the $\ell{=}1$ solution is degenerate under the Weyl group action,
\begin{equation}
(m,n) = (2,2), (2,0), (0,0) \qquad\Leftrightarrow\qquad [\pb,\qb]=[0,2]\ .
\end{equation}
With increasing level the real roots hug the lightcone,
as is apparent from Figure~4.
\begin{figure}[h!tbp]
\centering
\vspace{-3cm}
\includegraphics[width = 0.6\paperwidth]{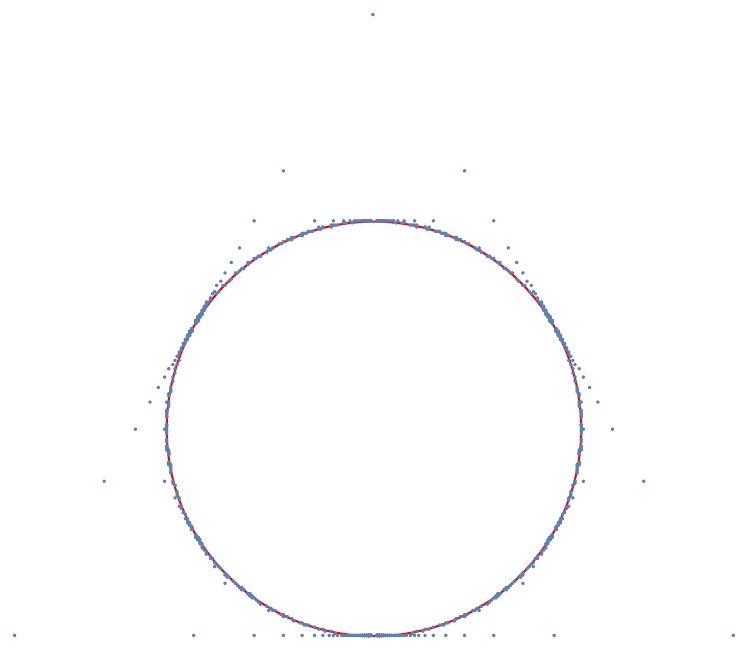} 
\caption{Projective view $\tfrac{\al}{|\al{\cdot}e_0|}-e_0$ of the real roots $\al$
for $1\le\ell\le100$ (the red circle is the lightcone)}
\end{figure}
The roots of the first family are also neatly expressed as
\begin{equation}
\al \= \ell\,n_i\pm\al_i \quad\for\quad i=1,2,3 
\qquad\with\quad\al_{\textrm{hw}}=\ell\,n_2+\al_2
\end{equation}
in terms of a triplet of null vectors
\begin{equation}\label{null}
\begin{aligned}
n_1 &\= \al_0-\al_2-\al_3 \= \sqrt{\tfrac23}\,e_0+\tfrac13(\al_3-\al_2)
\= \sqrt{2}\,\bigl( \tfrac{1}{\sqrt{3}}e_0-\tfrac{1}{\sqrt{3}}e_2\bigr)\ ,\\
n_2 &\= \al_0+\al_1-\al_3 \= \sqrt{\tfrac23}\,e_0+\tfrac13(\al_1-\al_3)
\= \sqrt{2}\,\bigl( \tfrac{1}{\sqrt{3}}e_0+\tfrac12 e_1+\tfrac{1}{2\sqrt{3}}e_2\bigr)\ ,\\
n_3 &\= \al_0+\al_1+\al_2 \= \sqrt{\tfrac23}\,e_0+\tfrac13(\al_2-\al_1)
\= \sqrt{2}\,\bigl( \tfrac{1}{\sqrt{3}}e_0-\tfrac12 e_1+\tfrac{1}{2\sqrt{3}}e_2\bigr)\ ,
\end{aligned}
\end{equation}
which satisfy the relations
\begin{equation}
n_i{\cdot}\,n_i=0\ ,\quad
n_i{\cdot}\,n_j=-1\ (i{\neq}j)\ ,\quad
\al_i{\cdot}\,n_j={\textstyle\sum}_k\epsilon_{ijk}\ ,\quad
n_1{+}n_2{+}n_3=\sqrt{6}\,e_0\ ,\quad
(\ell\,n_i\pm\al_i)^2=2\ .
\end{equation}

Since the real roots
\begin{equation}
\al\=\sqrt{2}\bigl(\sinh\eta\,e_0\ +\ \cosh\eta\cos\chi\,e_1\ +\ \cosh\eta\sin\chi\,e_2\bigr)
\end{equation}
are spacelike and located on a one-sheeted hyperboloid,
the fix planes of the corresponding reflections~$s_\al$,
\begin{equation}
\al{\cdot}x \= 0 \qquad\Leftrightarrow\qquad 
\tanh\eta \= \tanh\theta\,\cos(\chi{-}\phi)\ ,
\end{equation}
are timelike planes through the origin, which we call ``mirrors''.
They intersect the lightcone and the two-sheeted hyperboloid $x{\cdot}x=-1$.
The reflections~$s_\al$ act on the Minkowskian components~$x^\mu$
of a point~$x$ as $(s_\al x)^\mu = S^\mu_{\ \nu} x^\nu$ with
\begin{equation}
\bigl(S^\mu_{\ \nu}\bigr) \= 
\begin{pmatrix} 1 & 0 & 0 \\ 0 & \cos\chi & -\sin\chi \\ 0 & \sin\chi & \cos\chi \end{pmatrix}
\begin{pmatrix} \cosh 2\eta & -\sinh 2\eta & 0 \\ \sinh 2\eta & -\cosh 2\eta & 0 \\ 0 & 0 & 1 \end{pmatrix}
\begin{pmatrix} 1 & 0 & 0 \\ 0 & \cos\chi & \sin\chi \\ 0 & -\sin\chi & \cos\chi \end{pmatrix}\ .
\end{equation}
Each such hyperbolic reflection preserves the radial coordinate $r=\sqrt{-x{\cdot}x}$ 
and the time orientation but reverses the spatial orientation ($\det S=-1$), 
hence it represents an involution on the future hyperboloid.
A collection of such mirrors is displayed in Figure~5, 
and Figure~6 shows their intersections with the $x^0{=}1$ plane.
\begin{figure}[h!tbp]
\centering
\vspace{-1cm}
\includegraphics[width = 0.5\paperwidth]{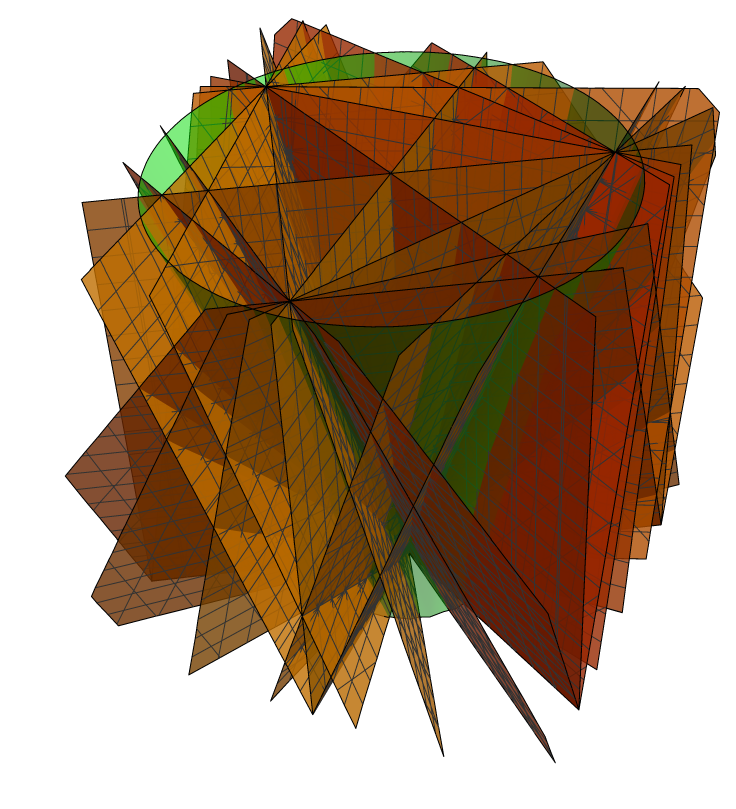} 
\vspace{-1cm}
\caption{The real-root mirrors for levels $|\ell|\le3$ (the green hyperboloid is $x{\cdot}x=2$)}
\end{figure}
\begin{figure}[h!tbp]
\centering
\includegraphics[width = 0.45\paperwidth]{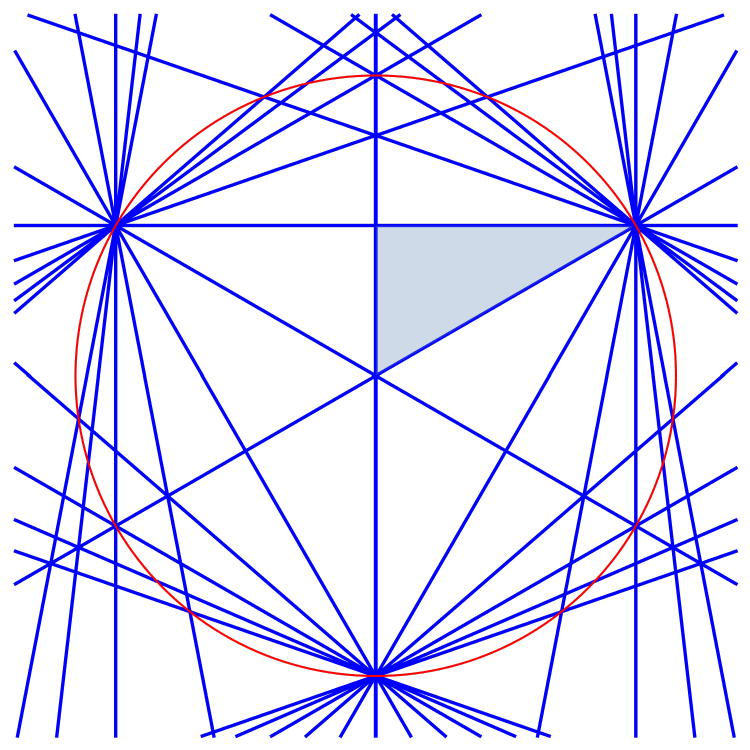} 
\caption{Intersection of the mirror planes (blue) for $|\ell|\le5$ and the lightcone (red) 
with the $x^0{=}1$ plane. The standard fundamental alcove is shaded.}
\end{figure}
\section{The potential}
\noindent
The ``horizontal'' $A_2$ slicing of the root space into levels~$\ell\in\Z$ leads to a decomposition
of the potential,
\begin{equation}
V(x) \= \sum_{\ell\in\Z} V_\ell(x) \qquad\with\quad 
V_\ell(x) \= \tfrac12\sum_{\al\in{\cal R}_\ell}\frac{1}{(\al\cdot x)^2}\ ,
\end{equation}
where ${\cal R}_\ell$ denotes the set of real roots~$\al$ with $\al{\cdot}e_0=\sqrt{\tfrac23}\,\ell$.
Clearly, $V_{-\ell}=V_\ell$.

Let us take a look at levels zero and one. Summing over the adjoint representation of~$A_2$,
\begin{equation}
V_0(r,\theta,\phi) \= \sum_{i=1,2,3} \frac1{(\al_i \cdot x)^2} \= \frac{\mathrm{csch}^{\!2}\theta}{2\,r^2}\biggl(
\frac1{\cos^2\!\phi} + \frac1{\cos^2 (\phi{-}\tfrac{2\pi}{3})} + \frac1{\cos^2 (\phi{+}\tfrac{2\pi}{3})} \biggr)
\= \frac9{2\,r^2}\,\frac{\mathrm{csch}^{\!2}\theta}{\cos^2\!3\phi}
\end{equation}
is just the celebrated P\"oschl--Teller potential, modulated by a $\theta$ dependence.
For level one we sum over the three extremal weights of the $[0,2]$ representation,
\begin{equation}
\begin{aligned}
&V_1(r,\theta,\phi) \= \sum_{i=1,2,3} \frac1{([n_i{+}\al_i] \cdot x)^2} \\
&\= \frac3{2\,r^2}\biggl(
\frac1{(\cosh\theta+2\sinh\theta\sin\phi)^2} + 
\frac1{(\cosh\theta+2\sinh\theta\sin(\phi{-}\tfrac{2\pi}{3}))^2} +
\frac1{(\cosh\theta+2\sinh\theta\sin(\phi{+}\tfrac{2\pi}{3}))^2}  \biggr) \\[4pt]
&\=\frac{18}{r^2}\,\frac{\cosh^4\!\theta+3\sinh^4\!\theta+4\cosh\theta\sinh^3\!\theta\sin 3\phi}
{(\cosh 3\theta-3\cosh\theta+4\sinh^3\!\theta\sin 3\phi)^2}\ .
\end{aligned}
\end{equation}

It is also possible to sum over whole families of solutions to the diophantine equation~(\ref{dio}).
Let us do so for the first family in~(\ref{series}), extending it to negative levels 
(to include the negative real roots) and including levels zero and one with their proper weight inside~$V$,
\begin{equation}
\begin{aligned}
V_{\textrm{1st family}} &\= -V_0-V_1+\ \sum_{\ell=0}^\infty\sum_{i=1,2,3}\biggl(
\frac1{([\ell\,n_i{+}\al_i] \cdot x)^2} + \frac1{([\ell\,n_i{-}\al_i] \cdot x)^2} \biggr)\\
&\= -V_0-V_1+\sum_{i=1,2,3} \frac2{(n_i{\cdot}x)^2}\ \sum_{\ell=0}^\infty
\frac{\ell^2 + \bigl(\tfrac{\al_i{\cdot}x}{n_i{\cdot}x}\bigr)^2}
{\bigl[ \ell^2 - \bigl(\tfrac{\al_i{\cdot}x}{n_i{\cdot}x}\bigr)^2 \bigr]^2}\\
&\=  -V_0-V_1+\sum_{i=1,2,3} \frac1{(n_i{\cdot}x)^2}\ \biggl( 
\frac1{\bigl(\tfrac{\al_i{\cdot}x}{n_i{\cdot}x}\bigr)^2} + 
\frac{\pi^2}{\sin^2\!\bigl(\pi\tfrac{\al_i{\cdot}x}{n_i{\cdot}x}\bigr)} \biggr)\\
&\= \sum_{i=1,2,3} \biggl(
\frac{\pi^2}{(n_i{\cdot}x)^2 \sin^2\!\bigl(\pi\tfrac{\al_i{\cdot}x}{n_i{\cdot}x}\bigr)}
\ -\ \frac1{([n_i{+}\al_i] \cdot x)^2} \biggr)\ ,
\end{aligned}
\end{equation}
where the $V_0$ contribution got cancelled on the way.
Although this is only a part of the full potential, it does show some characteristic features of~$V$:
\begin{itemize}
\item the $A_2$ subgroup's Weyl group $S_3$ yields a dihedral symmetry and six-fold mirrors,
\item infinitely many mirrors intersect in the three null lines $\lambda n_i$,
\item in the $\ell\to\infty$ limit the mirrors accumulate in the three null planes $n_i{\cdot}x=0$.
\item the mirrors tessalate the interior of the lightcone with infinitely many triangular Weyl alcoves
\item a fundamental Weyl alcove is spanned by $\{e_0, 2e_0{+}e_2, 2e_0{+}\sqrt{3}e_1{+}e_2\}$
\end{itemize}
A contour plot of $\log V$ on the plane $x^0{=}1$ is given in Figure~7 for one Weyl alcove.
\begin{figure}[h!tbp]
\centering
\includegraphics[width = 0.6\paperwidth]{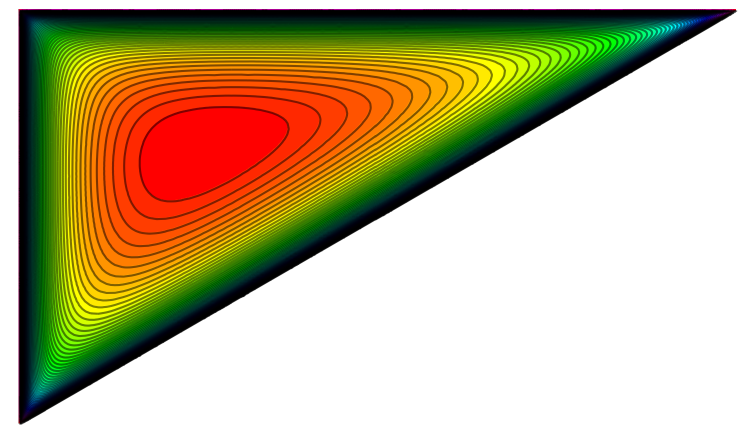} 
\caption{Contour lines of $\log V$ for the standard fundamental alcove intersecting the $x^0{=}1$ plane}
\end{figure}
\section{Mapping to the complex half-plane}
\noindent
Since our model is scale invariant, for the potential we can restrict ourselves to the future hyperboloid 
$r^2=-x{\cdot}x=1$ and $x^0\ge1$. It is convenient to pass to complex embedding coordinates
\begin{equation} \label{twdef}
t = x^0 \und w = x^1+\im x^2 \qquad\with\quad t^2-\bar{w}w = 1\ .
\end{equation}
By a stereographic projection (see Figure~8) the hyperboloid gets mapped 
to the unit disk $\bar{v}v\le1$ for $v\in\C$,
\begin{equation}
\frac{w}{t{+}1}=\frac{v}{1} \qquad\Rightarrow\qquad
w=\frac{2v}{1-\bar{v}v}\ ,\quad
v=\frac{w}{1+\sqrt{1{+}\bar{w}w}}\ ,\quad
t=\frac{1+\bar{v}v}{1-\bar{v}v}\ .
\end{equation}
\begin{figure}[h!tbp]
\centering
\includegraphics[width = 0.6\paperwidth]{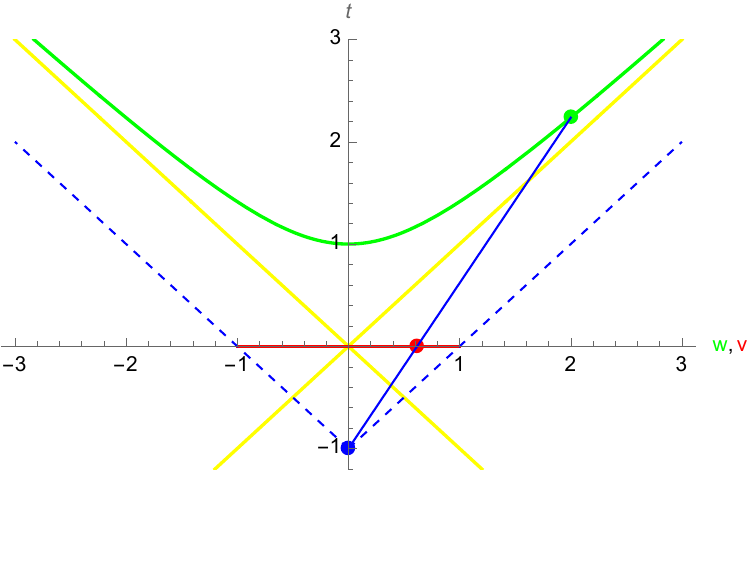} 
\vspace{-1.5cm}
\caption{Stereographic projection of the hyperboloid (green) to the disk (red),
with the lightcone (yellow)}
\end{figure}
The metric induced from the Minkowski metric turns this into the Poincar\'e disk model 
of the hyperbolic plane~$H^2$.
The intersection curves with the mirrors of level zero and one are easily computed as
\begin{equation} \label{vcurves}
\begin{aligned}
\ell=0:&\qquad\, v+\bar{v}=0\ ,\qquad\quad\
\bar{\rho}\,v+\rho\,\bar{v}=0\ ,\qquad\
\rho\,v+\bar{\rho}\,\bar{v}=0\ ,\\
\ell=1:&\qquad |v-2\im|^2 = 3\ ,\qquad
|v-2\im\rho|^2 = 3\ ,\qquad
|v-2\im\bar{\rho}|^2= 3\ ,
\end{aligned}
\end{equation}
where
\begin{equation}
\rho = \ep^{2\pi\im/3} \qquad\Rightarrow\qquad
\rho^2=\bar\rho\ ,\quad \rho+\bar\rho=-1\ ,\quad 1+\rho+\rho^2=0\ .
\end{equation}
Adding the infinity of real-root mirrors produces a paracompact triangular tessalation 
of type $[p,q,r]=[2,3,\infty]$. Each of the hyperbolically congruent triangles has angles
$\frac{\pi}{2}$, $\frac{\pi}{3}$ and~$0$, and at the corresponding vertices there meet
4, 6 and infinitely many triangles, thus one vertex is always at the boundary,
as is visible from Figure~9.
\begin{figure}[h!tbp]
\centering
\includegraphics[width = 0.45\paperwidth]{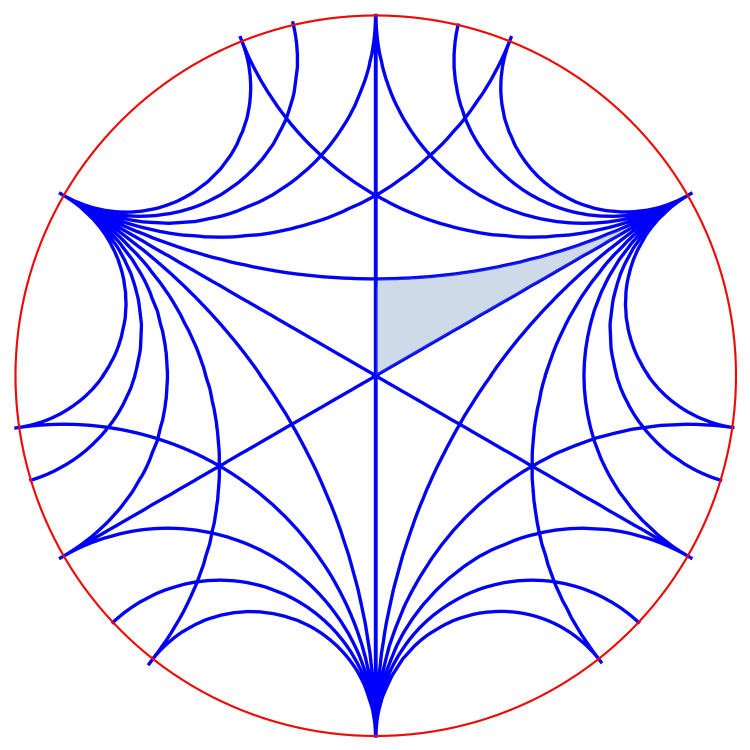} 
\caption{The mirror lines (blue) for $|\ell|\le5$ in the Poincar\'e disk $\bar{v}v{<}1$
(with red boundary $\bar{v}v{=}1$)}
\end{figure}

We employ a variant of the Cayley map to further pass to the complex upper half plane~${\cal H}\ni z$,
\begin{equation}
v = -\im\,\frac{\rho\,z+1}{z+\rho} \qquad\Leftrightarrow\qquad
z = \frac{1-\im\rho\,v}{\im\,v-\rho}\ ,
\end{equation}
such that the boundary $|v|{=}1$ becomes the real axis $\textrm{Im}\,z{=}0$. The direct relation between $w$ and~$z$ reads
\begin{equation} \label{wtoz}
w \= \frac{2}{\sqrt{3}}\,\frac{(1{+}\rho\,z)(\bar{\rho}{+}\bar{z})}{z-\bar{z}}
\= \frac{2}{\sqrt{3}}\,\frac{\rho\,z\bar{z}+(z{+}\bar{z})+\bar{\rho}}{z-\bar{z}}
\= \frac{2\,\rho}{\sqrt{3}}\,\frac{(z{+}\bar{\rho})(\bar{z}{+}\bar{\rho})}{z-\bar{z}}\ ,
\end{equation}
and the mirror curves at level zero and one become (in the same order as in (\ref{vcurves}))
\begin{equation} \label{zcurves}
\begin{aligned}
\ell=0:&\qquad z\,\zb=1\ ,\qquad\quad\ z+\zb=1\ ,\qquad\qquad\quad\ (z{-}1)(\zb{-}1)=1\ ,\\
\ell=1:&\qquad z+\zb=0\ ,\qquad (2z{-}1)(2\zb{-}1)=1\ ,\qquad z+\zb=2\ .
\end{aligned}
\end{equation}
The first two curves in the $\ell{=}0$ list and the first one at $\ell{=}1$ bound a standard
fundamental domain
\begin{equation} \label{funddom}
{\cal F} \= \bigl\{ z\in{\cal H} \ \bigm| \ |z|\ge1\und 0\le\Re\,z\le\tfrac12 \bigr\}\ ,
\end{equation}
which is co-finite (with a hyperbolic volume of~$\tfrac{\pi}{6}$) but not co-compact due to a cusp at~$\im\infty$.
Figure~10 shows the mirror lines of Figure~9 mapped to the upper half plane~$\cal H$.
\begin{figure}[h!tbp]
\centering
\includegraphics[width = 0.6\paperwidth]{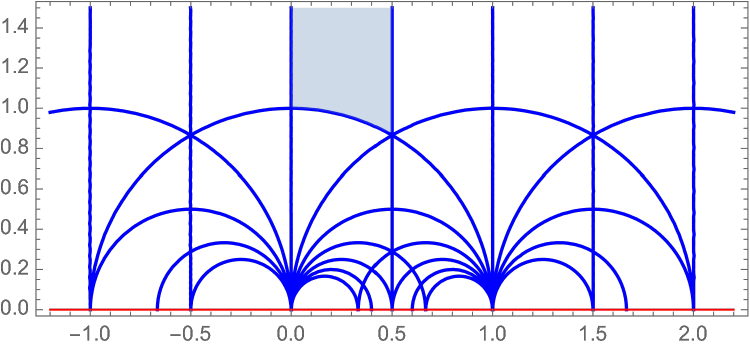} 
\caption{The mirror lines (blue) for $|\ell|\le5$ in the upper half plane $\Im\,z>0$
(with red boundary $\Im\,z=0$)}
\end{figure}
Any other triangle in the tessalation is reached by applying a suitable element of PGL($2,\Z$),
the group of integral $2{\times}2$ matrices with determinant $+1$ or~$-1$ modulo~$\{\pm\unity\}$:
\begin{equation}
z\ \mapsto\ \begin{cases}
\ \tfrac{az+b}{cz+d}       & \textrm{if}\quad \bigl|\begin{smallmatrix} a&b\\c&d \end{smallmatrix}\bigr|=+1 \\[4pt]
\ \tfrac{a\zb+b}{c\zb+d} & \textrm{if}\quad \bigl|\begin{smallmatrix} a&b\\c&d \end{smallmatrix}\bigr|=-1 
\end{cases}
\qquad\with a,b,c,d\in\Z\ .
\end{equation}
This happens to be the Weyl group of our hyperbolic Kac-Moody algebra.
It can be generated by the three reflections
\begin{equation}
s_1=s_{\al_1}:\quad z\mapsto 1/\zb\ ,\qquad
s_2=s_{\al_2}:\quad z\mapsto 1{-}\zb\ ,\qquad
s_3=s_{\al_0-2\al_3}:\quad z\mapsto -\zb\ ,
\end{equation}
whose fixpoints form the three mirror curves mentioned above, 
which bound the fundamental triangle~(\ref{funddom}).
The two generators of the even subgroup PSL($2,\Z$) are
\begin{equation}
T = s_2 s_3:\quad z\mapsto z{+}1 \quad\und\quad
S = s_1 s_3:\quad z\mapsto -1/z\ ,
\end{equation}
and its standard fundamental domain is cut in half by the extra reflection~$s_3$.
In matrix representation $(\begin{smallmatrix}a&b\\c&d\end{smallmatrix})$ we have
\begin{equation}
s_1 \hateq (\begin{smallmatrix} 0&1\\1&0 \end{smallmatrix}) \ ,\qquad
s_2 \hateq (\begin{smallmatrix} 1&-1\\0&-1 \end{smallmatrix}) \ ,\qquad
s_3 \hateq (\begin{smallmatrix} 1&0\\0&-1 \end{smallmatrix}) \ ,\qquad
T \hateq (\begin{smallmatrix} 1&1\\0&1 \end{smallmatrix}) \ ,\qquad
S \hateq (\begin{smallmatrix} 0&-1\\1&0 \end{smallmatrix}) \ ,
\end{equation}
up to multiplication by $-1$ of course.
The simple-root reflection $s_{\al_0}: z\mapsto\frac{\zb}{2\zb-1}$ appears
in the middle of our $\ell{=}1$ lists (\ref{vcurves}) and~(\ref{zcurves}).
Choosing it instead of $s_3$ leads to a fundamental domain with the cusp
sitting at $1$ rather than~$\im\infty$.
In any case, it is clear that the potential 
$U(z)=V\bigl(r{=}1,\theta(z),\phi(z)\bigr)$
is a real automprphic function with respect to PGL($2,\Z$).
We end this section by displaying $\log U$ in the Poincar\'e disk and 
in the upper half plane for the standard fundamental domain in Figure~11.
\begin{figure}[h!tbp]
\centering
\includegraphics[width = 0.48\paperwidth]{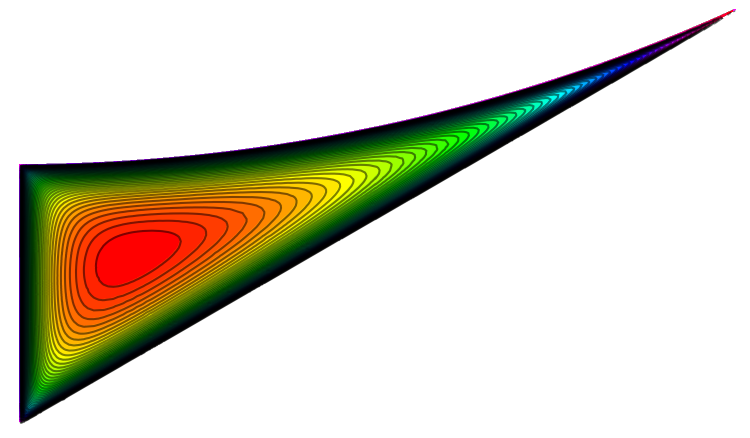}  \qquad
\includegraphics[width = 0.22\paperwidth]{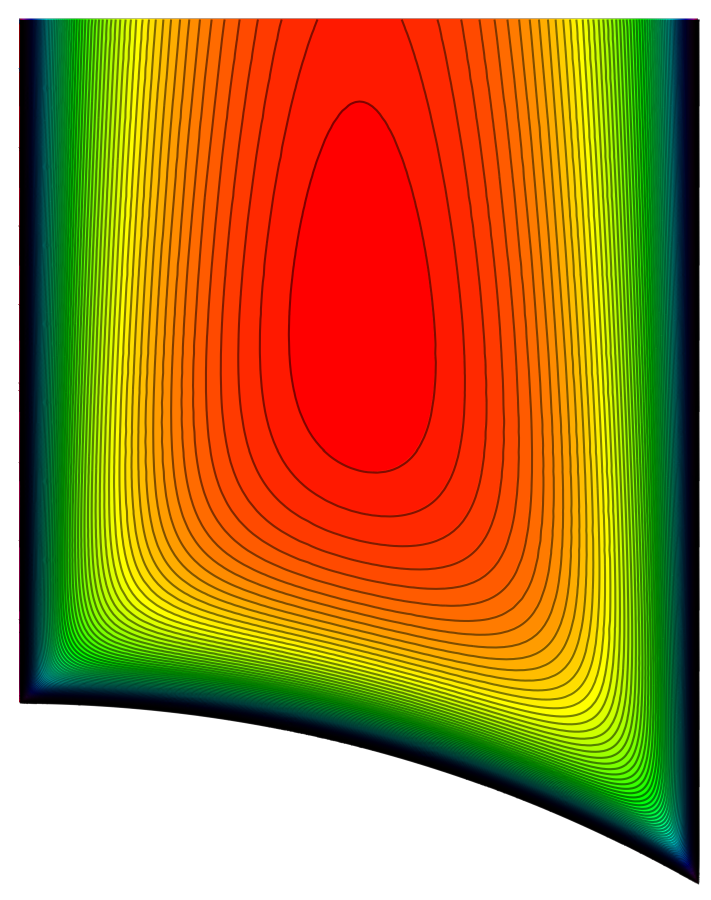}  
\caption{Contour lines of $\log U$ for a fundamental domain
in the $v$ disk (left) and the $z$ plane (right)}
\end{figure}
%

%
\section{The potential as a Poincar\'e series}
\noindent
Our potential is a sum over all real roots~$\al$ of $AE_3$,
thus each reflection~$s_\al$ inside PGL($2,\Z$) provides one summand.
These reflections are given by traceless matrices~$R$ of determinant~$-1$,\footnote{
The matrix entries $(p,q,r)$ are not to be confused with the tessalation labels.}
\begin{equation} \label{reflectz}
s_\al \equiv s_R\ :\quad
z\ \mapsto\ \frac{p\zb+q}{r\zb-p} \qquad\with\quad 
R = \begin{pmatrix} p & q \\[2pt] r & -p \end{pmatrix}
\quad\und\quad p^2+q\,r=1\ .
\end{equation}
The scalar product in root space is an invariant bilinear form,
\begin{equation}
\al{\cdot}\al' \= \tr ( s_\al\,s_{\al'} ) \= 2\,p\,p'+q\,r'+r\,q'\ \in \Z\ .
\end{equation}
From (\ref{wtoz}) we see that the real function $\al{\cdot}x$ odd under $s_\al$
becomes a real quadratic polynomial in $z$ and~$\zb$ divided by $|z{-}\zb|$.
A quick computation shows that 
\begin{equation} \label{mirrorpqr}
\al{\cdot}x \= \tfrac{\sqrt{2}\,\im}{z - \zb}\,\bigl[r\,z\,\zb - p(z{+}\zb) - q \bigr]
\end{equation}
is indeed odd under the reflection~(\ref{reflectz}). 
Therefore, on the upper half plane the potential is expressed as
\begin{equation} \label{UR}
U(z) \equiv V\bigl(x(z)\bigr) 
\= \frac14 \sum_R \frac{|z-\zb|^2}{[r\,z\,\zb - p(z{+}\zb) - q]^2}
\= \frac14 \sum_R \frac{4\,y^2}{[r(x^2{+}y^2)-2p\,x-q]^2}
\ =:\ \frac14 \sum_R u_R(z)
\end{equation}
where $z=x+\im\,y$ and the sum runs over all reflections~$R$ in~(\ref{reflectz}),
with $R$ and~$-R$ contributing the same.
Comparison with
\begin{equation} \label{mirrorlmn}
-\al{\cdot}x \=  \tfrac{\sqrt{2}\,\im}{z - \zb}\,
\bigl[ (2\ell{-}m)\,z\zb + (n{-}\ell)(z{+}\zb) + (m{-}n)\bigr]
\end{equation}
provides the translation between the labels $(p,q,r)$ and $(\ell,m,n)$, 
up to a common sign of course.
As an aside, we characterize in Table~1
the two orbits (e for $\ell$ even, o for $\ell$ odd) 
of the Weyl group on the set $\cal R$ of real roots.
\begin{table}[htbp]
\begin{center}
\begin{tabular}{|c|ccc|ccc|}
\hline
& $\ell$ & $m$ & $n$ & $p$ & $q$ & $r$ \\
\hline
& e & o & o & o & e & o \\
${\cal R}_+$ & e & e & o & o & o & e \\
& e & o & e & e & o& o \\
\hline
${\cal R}_-$ & o & e & e & o & e & e \\
\hline
\end{tabular}
\caption{Assignment of Weyl orbit to a root $\al(\ell,m,n)$ or to a reflection $R(p,q,r)$ (e = even, o = odd)}
\end{center}
\vspace{-5mm}
\end{table}
Restricting our potential to one of the two Weyl orbits removes either the $\ell{=}\textrm{odd}$
or the $\ell{=}\textrm{even}$ mirror lines from all diagrams and doubles the fundamental domain.
In an Appendix with Don Zagier we outline how far one can proceed 
with an explicit computation of the potential function~$U(z)$.

The potential~$U(z)$ is a real modular function under the action 
of GL($2,\Z$), which is manifest via
\begin{equation}
u_R (M z) \= u_{M^{-1}R\,M}(z) \qquad\for\quad M z=\tfrac{az+b}{cz+d}
\quad\with\quad a,b,c,d\in\Z\ .
\end{equation}
Hence, we can replace the sum over~$R$ with sums over appropriate orbits by
the adjoint action of~GL($2,\Z$) for a suitable reference, say
\begin{equation}
R_0 \=  \bigl(\begin{smallmatrix} 1 & 0 \\[2pt] 0 & -1 \end{smallmatrix}\bigr)
\qquad\Leftrightarrow\qquad z\ \mapsto\ -\zb
\qquad\Leftrightarrow\qquad u_{R_0}(z) = \frac{y^2}{x^2} =: u(z)\ ,
\end{equation}
and obtain~\footnote{
The left-hand side is doubled because    
$\{c,d,-a,-b\}$ yields the same adjoint action as $\{a,b,c,d\}$.}
\begin{equation}
2\,U(z) \= \frac14\sum_M u_{M^{-1}R_0 M}(z) \= \frac14\sum_M u(Mz)
\= \frac14\sum_M \frac{(ad{-}bc)^2\ y^2}{[ac(x^2{+}y^2)+(ad{+}bc)x+bd]^2}\ .
\end{equation}
The question is: over which subset of GL($2,\Z$) matrices does this sum run?
Since our fundamental domain~${\cal F}$ in (\ref{funddom}) is half of the standard one 
for~PSL($2,\Z$), all reflections~$R$ should be covered using adjoint orbits by matrices~$M_n$ 
of determinant $n=ad{-}bc$ equal to one or two. 
Indeed, for $R\in{\cal R}_+$, we can find a unique (up to sign and the footnote) matrix~$M_2$
such that $R=M_2^{-1}R_0M_2$. In the case of $R\in{\cal R}_-$, in constrast,
there are {\sl two\/} such matrices~$M_2$, which have the form
\begin{equation}
\Bigl(\begin{smallmatrix} 2a' & 2b' \\[4pt] c' & d' \end{smallmatrix}\Bigr) \quad\und\quad
\Bigl(\begin{smallmatrix} a' & b' \\[4pt] 2c' & 2d' \end{smallmatrix}\Bigr) \qquad\Rightarrow\qquad
\det\Bigl(\begin{smallmatrix} a' & b' \\[4pt] c' & d' \end{smallmatrix}\Bigr) = 1\ ,
\end{equation}
so such reflections~$R$ are covered twice by summing over~$M_2$. 
However, they also make up (again uniquely) the $M_1$~orbit of~$R_0$.
Therefore, we can correct the overcount by subtracting,
\begin{equation}
2\,U(z) \= F_2(z) - F_1(z) \qquad\for\qquad
F_n(z) \ :=\ \tfrac14 \sum_{M_n} u(Mz)\ .
\end{equation}

The function $F_2$ can be obtained from $F_1$ by applying a Hecke operator~$T_2$ 
(for weight $k{=}0$),
\begin{equation} \label{T2}
F_2(z) \= (T_2 F_1)(z) 
\= \sum_{\genfrac{}{}{0pt}{2}{ad=2}{a,d>0}} \sum_{b\,(\textrm{mod}\,d)} F_1\bigl(\tfrac{az+b}{d}\bigr)
\= F_1(2z) + F_1(\tfrac{z}{2}) + F_1(\tfrac{z+1}{2})\ ,
\end{equation}
and thus we have
\begin{equation} \label{UviaF}
2\,U(z) \= F_1(2z) + F_1(\tfrac{z}{2}) + F_1(\tfrac{z+1}{2}) - F_1(z)\ .
\end{equation}
Therefore, it suffices to compute the Poincar\'e series
\begin{equation}
F_1(z) \= \tfrac14 \!\!\!\!\!\! \sum_{M\in\textrm{SL}(2,\Z)} \!\!\!\! u(Mz)\ .
\end{equation}

There is another path to this result, which provides a useful connection to binary quadratic forms.\footnote{
I thank Don Zagier for pointing out this and~(\ref{T2}).}
Let us define~\cite{Zag24}
\begin{equation}
\tF_D(z) \= \frac12 \sum_{Q(D)} \frac{D\ y^2}{[A(x^2{+}y^2)+B\,x+C]^2}
\= \frac12 \sum_{Q(D)} \frac{D\ y^2}{|A\,z^2+B\,z+C|^2 - y^2}\ ,
\end{equation}
where $Q(D)$ denotes the set of binary quadratic forms $As^2{+}Bst{+}Ct^2$ over~$\Z$
with discriminant~$D=B^2{-}4AC$.
For $D{=}1$ there is a bijection between $Q(1)$ and PSL($2,\Z$) by solving
\begin{equation}
a\,c=A\ ,\qquad b\,d = C\ ,\qquad a\,d = \sfrac12(B{+}1)\ ,\qquad b\,c = \sfrac12(B{-}1)\ .
\end{equation}
We conclude that $\tF_1=F_1$. 
For $D{=}4$ we have a bijection between $Q(4)$ and our reflections~$R$ in~(\ref{reflectz}),
\begin{equation}
r = A\ ,\qquad -2p = B\ ,\qquad -q = C\ .
\end{equation}
Therefore, the potential can also be expressed as
\begin{equation} \label{F4}
2\,U(z) \= \tF_4(z) \= \frac12 \sum_{Q(4)} \frac{4\ y^2}{[A(x^2{+}y^2)+B\,x+C]^2}\ .
\end{equation}
Now, $Q(4)$ can be reduced to $Q(1)$, and it is not too hard to check that indeed~\cite{Zag24}
\begin{equation}
\tF_4(z) \= (T_2 \tF_1)(z) - \tF_1(z) \= \tF_1(2z) + \tF_1(\tfrac{z}{2}) + \tF_1(\tfrac{z+1}{2}) - \tF_1(z) \ ,
\end{equation}
confirming~(\ref{UviaF}).
It thus suffices to compute the generalized real-analytic Eisenstein series
\begin{equation} \label{Poincare}
\tF_1(z) \= \frac12 \sum_{Q(1)} \frac{y^2}{[A(x^2{+}y^2)+B\,x+C]^2}
\= \frac12 \sum_{Q(1)} \frac{y^2}{|A\,z^2+B\,z+C|^2 - y^2}\ ,
\end{equation}
where $Q(1)$ indicates a discriminant $B^2-4AC=1$.
As was shown in~\cite{Zag24}, this sum converges almost everywhere.\footnote{
On the mirror curves one summand is infinite.
The critical part is the $A{=}0$ subsum, $\sum_{C\in\Z} y^2(x{+}C)^{-2}=\pi^2 y^2\sin^{-2}\!\pi x$.}
However, it does not decay at~$\im\infty$, but grows as $\tF_1(x{+}\im y)\sim y^2$ for $y\to\infty$.

The form (\ref{F4}) can be translated back to the unit hyperboloid and indeed the Minkowski future,
with the result
\begin{equation}
V(x) \= \sum_{Q(4)} \frac{3}{\bigl[ 
(2A{-}B{+}2C)\,x^0 + \sqrt{3}(A{-}C)\,x^1 - (A{-}2B{+}C)\,x^2\bigr]^2}\ ,
\end{equation}
where the sum runs over all integers $A$, $B$ and $C$ subject to the $Q(4)$ condition
$B^2-4AC=4$. In this way, the real roots are parametrized by binary quadratic forms,
which is of course equivalent to the solutions of the diophantine equation~(\ref{dio})
but may be more convenient or manageable.
\section{Dunkl operators}
\noindent
Calogero models in a Euclidean space are known to be maximally superintegrable. 
This is also the case for the spherical reduction of the rational models. 
One key instrument to establish this property is the linear Dunkl operators~\cite{Dunkl, Heckman}
\begin{equation} \label{Dunkl}
{\cal D}_i \= \pa_i\ -\ \tfrac{g}{2}\sum_{\al\in\cal R} \frac{\al_i}{\al{\cdot}x}\,s_\al
\qquad\textrm{for}\quad i=1,\ldots,n
\end{equation}
and their angular versions
\begin{equation}
{\cal L}_{ij} \= x^i {\cal D}_j - x^j {\cal D}_i\ ,
\end{equation}
respectively. 
For rational models, the crucial property is the commutation $[{\cal D}_i,{\cal D}_j]=0$,
while the ${\cal L}_{ij}$ deform the angular momentum algebra to
a subalgebra of a rational Cherednik algebra~\cite{FeiHak}.
It is known that every Weyl-invariant polynomial in the ${\cal D}_i$
or in the ${\cal L}_{ij}$ will, upon its restriction `res' to Weyl-invariant functions, 
provide a conserved quantity, {\em i.e.}~an operator which commutes
with the Hamiltonian $H$ or $H_\Omega$, respectively.
Indeed, the Hamiltonians themselves can be expressed in this way,
\begin{equation}
\begin{aligned}
-2\,H &\= \textrm{res} \sum_{i=1}^n {\cal D}_i^2 
\= \sum_{i=1}^n \pa_i^2 - \textrm{res} \sum_{\al\in{\cal R}} 
\frac{\al{\cdot}\al}{2(\al{\cdot}x)^2}\,g(g{-}s_\al)\ ,\\[4pt]
-2\,H_\Omega &\= \textrm{res} \sum_{i<j} {\cal L}_{ij}^2 - 2E_0 
\= L^2 - \textrm{res} \sum_{\al\in{\cal R}} 
\frac{\al{\cdot}\al\,x{\cdot}x}{2(\al{\cdot}x)^2}\,g(g{-}s_\al) \ ,
\end{aligned}
\end{equation}
with the ground-state energy 
\begin{equation}
E_0 \= \tfrac12\,\textrm{res}\,gS(gS{+}n{-}2) \qquad\textrm{for}\quad 
S= \tfrac12{\textstyle\sum}_\al s_\al\ .
\end{equation}

Let us repeat this construction for $\R^{1,2}$ and the restriction to the hyperboloid $x{\cdot}x=-r^2=-1$.
We follow the standard construction and define the `hyperbolic Dunkl operators' ($i=1,2$)
\begin{equation}
{\cal C} \= x^1\pa_2{-}x^2\pa_1 - \tfrac{g}{2}\sum_{\al\in\cal R}\sfrac{x^1\al^2{-}x^2\al^1}{\al{\cdot}x}\,s_\al
\quad\und\quad
{\cal B}_i \= -x^0\pa_i-x^i\pa_0 + \tfrac{g}{2}\sum_{\al\in\cal R}\sfrac{x^0\al^i{-}x^i\al^0}{\al{\cdot}x}\,s_\al
\end{equation}
with $\al{\cdot}x=-\al^0 x^0+\al^i x^i$ as deformed rotation and boost generators.
Lorentz indices are raised and lowered with the Minkowski metric.
In complex coordinates~(\ref{twdef}) on the hyperboloid these Dunkl operators read~\footnote{
The differential parts are $C = \pa_\phi$ and $B_\pm = \ep^{\pm\im\phi}(\pa_\theta\pm\im\coth\theta\pa_\phi)$.}
\begin{equation} \label{hypDunkl}
\begin{aligned}
{\cal C} &\= \im(w\pa_w{-}\wb\pa_\wb) 
- \im\,\tfrac{g}{4} \sum_{\al\in\cal R}\tfrac{w\al^\wb{-}\wb\al^w}{\al{\cdot}x}\,s_\al \quad\und \\
{\cal B}_+ &\= {\cal B}_1{+}\im{\cal B}_2 \= 2\sqrt{1{+}w\wb}\;\pa_\wb 
+\tfrac{g}{2}\sum_{\al\in\cal R}\tfrac{t\al^w{-}w\al^t}{\al{\cdot}x}\,s_\al \= ({\cal B}_-)^*\ .
\end{aligned}
\end{equation}
In half-plane coordinates they take the form
\begin{equation}
\begin{aligned}
\sqrt{3}\;{\cal C} \ \;&\= (z{+}\rho)(z{+}\bar{\rho})\pa_z + (\zb{+}\rho)(\zb{+}\bar{\rho})\pa_\zb
\ +\ \tfrac{g}{2}\sum_{\al\in{\cal R}} 
\frac{(r{-}2p)z\zb-(q{+}r)(z{+}\zb)+(q{+}2p)}{r\,z\zb-p\,(z{+}\zb)-q} \, s_R \ , \\
\sqrt{3}\;{\cal B}_+ &\= \rho\,(z{+}\bar{\rho})^2\pa_z+\rho\,(\zb{+}\bar{\rho})^2\pa_\zb
\ +\ \tfrac{g}{2}\sum_{\al\in{\cal R}} 
\frac{(2r{+}2\rho p)z\zb+(\rho q{+}\bar{\rho}r)(z{+}\zb)+(2q{-}2\bar{\rho}p)}{r\,z\zb-p\,(z{+}\zb)-q} \, s_R \ , \\
\sqrt{3}\;{\cal B}_- &\= \bar{\rho}\,(z{+}\rho)^2\pa_z+\bar{\rho}\,(\zb{+}\rho)^2\pa_\zb 
\ +\ \tfrac{g}{2}\sum_{\al\in{\cal R}} 
\frac{(2r{+}2\bar{\rho}p)z\zb+(\bar{\rho}q{+}\rho r)(z{+}\zb)+(2q{-}2\rho p)}{r\,z\zb-p\,(z{+}\zb)-q} \, s_R \ .
\end{aligned}
\end{equation}
These operators obey the algebra
\begin{equation}
[{\cal C}\,,\, {\cal B}_\pm] \= \pm\im\,{\cal B}_\pm \ +\ O(g)\qquad\und\qquad
[{\cal B}_+\,,\,{\cal B}_-] \= -2\im\,{\cal C} \ +\ O(g)\ ,
\end{equation}
which at $g{=}0$ reduces to the $so(1,2)$ algebra.
The $O(g)$ deformations are determined by the action of the differential parts
on the reflection parts and by the commutators of the reflection parts themselves.

A standard computation shows that the commutator of two linear Dunkl operators, 
$[{\cal D}_\mu,{\cal D}_\nu]$,
reduces to the antisymmetric part (under $\mu\leftrightarrow\nu$) of
\begin{equation}
g^2 Y_{\mu\nu} \ :=\ \tfrac{g^2}{4} {\sum_{\al,\beta}}^\prime 
\frac{\al_\mu\beta_\nu}{\al{\cdot}x\ s_\al(\beta{\cdot}x)}\,s_\al s_\beta 
\= \ \tfrac{g^2}{4} {\sum_{\al,\beta}}^\prime 
\frac{\al_\mu\beta_\nu-(\al{\cdot}\beta)\al_\mu \al_\nu}{\al{\cdot}x\ \beta{\cdot}x}\,s_\beta s_\al\ ,
\end{equation}
where the prime indicates excluding pairs with $\beta{=}{\pm}\al$.
In the last step, under the sum we substituted $\beta\to s_\al\beta=\beta-(\beta{\cdot}\al)\,\al$, 
{\em i.e.}~$s_\beta\to s_\al s_\beta s_\al$, and used $s_\al\al=-\al$ or $s_\al^2=1$.
Hence, the criterion for linear Dunkl operators to commute is the vanishing of a two-form,
\begin{equation} \label{Yform}
Y\ \equiv\ Y_{\mu\nu}\,\diff x^\mu{\wedge}\diff x^\nu \=
\tfrac18 {\sum_{\al,\beta}}^\prime 
\frac{\al\wedge\beta}{\al{\cdot}x\ \beta{\cdot}x}\,[s_\al,s_\beta] 
\ \buildrel{!}\over{=}\ 0
\end{equation}
where we abbreviated $\al_\mu\diff x^\mu{=}\al$ and $\beta_\nu \diff x^\nu{=}\beta$.
Note that the four pairs $(\al,\pm\beta)$ and $(-\al,\pm\beta)$ contribute equally to the double sum.

In order to generate the angular Hamiltonian $H_\Omega$ in~(\ref{hypHam}), we compute
\begin{equation} \label{L2hyp}
\sum_{\mu<\nu}{\cal L}_{\mu\nu}{\cal L}^{\mu\nu} \=
{\cal C}^2 - \tfrac12({\cal B}_+{\cal B}_- + {\cal B}_-{\cal B}_+)
\= -L^2 + \sum_{\al\in{\cal R}} \frac{x{\cdot}x}{(\al{\cdot}x)^2}\,g(g{-}s_\al) 
+ gS(gS{+}1) - g^2\,\eta^{\mu\nu}Y_{\mu\nu}
\end{equation}
by generalizing the results in \cite{FeiHak} to $\R^{1,2}$.
We remark that, due to the indefinite root-space signature and $x{\cdot}x{=}-1$, 
the relative sign between $\sum{\cal L}^2$ and the angular Hamiltonian is flipped and 
the ground-state energy $E_0=-\tfrac12\,\textrm{res}\,gS(gS{+}1)$ is negative and formally infinite.
Besides this energy shift, our hyperbolic Dunkl operators can generate the 
angular Hamiltonian provided that
$(Y_{\mu\nu})$ is not only symmetric but also traceless, {\em i.e.}
\begin{equation} \label{Ytrace}
\eta^{\mu\nu}Y_{\mu\nu} \= -\tfrac18 {\sum_{\al,\beta}}^\prime 
\frac{\al\cdot\beta}{\al{\cdot}x\ \beta{\cdot}x}\,\{s_\al,s_\beta\} 
\ \buildrel{!}\over{=}\ 0\ .
\end{equation}
\section{Integrability?}
\noindent
The commutativity of the linear Dunkl operators of rational Calogero models
is a key property for their integrability in Euclidean space.
It also assures the integrability of the angular models constructed by reduction to the sphere.
It is therefore reasonable to perform this test also for
our hyperbolic Kac--Moody Calogero model.\footnote{
To be sure, this is at best a sufficient condition. Already for Sutherland or elliptic models,
the linear Dunkl commutators are not vanishing but still take a simple form.}
We shall now investigate the conditions (\ref{Yform}) and~(\ref{Ytrace}), 
{\em i.e.}~$Y=0$ and $Y^\mu_{\ \mu}=0$. 

For classical root systems indeed $Y=0$, 
because the double sum in~(\ref{Yform}) can be recast as a sum over planes of contributions 
stemming from the real root pairs lying in a given plane~$\Pi$, which add up to zero for any such plane. 
In our hyperbolic model, this is obvious only for root pairs $(\al,\beta)$ at level~$\ell{=}0$,
which form the $A_2$ subalgebra with a hexagon of roots and $\al{\cdot}\beta{=}{\pm}1$ throughout.
Generically however, two arbitrary real roots $\al$ and $\beta$ generate 
an infinite planar collection of real roots,
\begin{equation} \label{planarsequence}
\al \longrightarrow s_\beta\al \longrightarrow s_\al s_\beta\al 
\longrightarrow s_\beta s_\al s_\beta \al \longrightarrow \ldots 
\quad\und\quad
\beta\longrightarrow s_\al\beta \longrightarrow s_\beta s_\al\beta 
\longrightarrow s_\al s_\beta s_\al \beta \longrightarrow \ldots
\end{equation}
and their negatives. 
The roots in either string are related by hyperbolic reflections and rotations,
but $\al$ and $\beta$ need not be.
All these comprise the real roots of a rank-2 subalgebra whose Cartan matrix reads~\cite{FN}
\begin{equation}
A_m \= \begin{pmatrix} \ph 2 & -m \\ -m & \ph 2 \end{pmatrix}
\qquad\for m=|\al{\cdot}\beta|\ \in\,\{0,1,2,3,4,\ldots\}
\end{equation}
and whose Weyl group is 
\begin{equation} \label{Weyl}
\bigl\{ (s_\al s_\beta)^{k-1}\ ,\ (s_\al s_\beta)^{k-1} s_\al \bigr\}
\qquad\for k\in\Z
\end{equation}
because $(s_\al s_\beta)^{-1}=s_\beta s_\al$ and $s_\al^2=s_\beta^2=\unity$.
Each odd element is a reflection on a hyperplane orthogonal to some real root~$\gamma_k$,
while the even elements are elliptic, parabolic or hyperbolic elements of PSL$(2,\Z)$,
for $m{\le}1$, $m{=}2$ or $m{\ge}3$, respectively.

Without loss of generality we can choose the signs of $\al$ and $\beta$ such that $\al{\cdot}\beta=-m\le0$.
Any real root in $\Pi_m=\langle\al\beta\rangle$ is a linear combination
\begin{equation} \label{quadric}
\gamma \= \xi\,\al\ +\ \eta\,\beta \qquad\with\quad (\xi,\eta)\in \Z^2 \und \xi^2+\eta^2-m\,\xi\,\eta= 1\ .
\end{equation}
Rather than finding the integral points on this quadric, we may compute the coefficients $\xi$ and~$\eta$
recursively from~(\ref{planarsequence}). 
We recombine these two sequences in an alternating fashion (and flipping half of the signs) 
into a double-infinite sequence
\begin{equation}
\gamma_{2\ell-1}=(s_\al s_\beta)^{\ell-1} \al \und \gamma_{2\ell}=(s_\al s_\beta)^{\ell-1} s_\al \beta
\qquad\Leftrightarrow\qquad s_{\gamma_k} = (s_\al s_\beta)^{k-1} s_\al
\end{equation}
with $k,\ell\in\Z$. Due to $s_\al \al=-\al$ and $s_\beta \beta=-\beta$ it reads
\begin{equation}
\bigl\{ \gamma_k \bigr\} \= \bigl\{ \ \ldots\ ,\ 
-s_\beta s_\al s_\beta \al\ ,\ -s_\beta s_\al \beta\ ,\ -s_\beta \al, -\beta\ ,\
 \al\ ,\ s_\al \beta\ ,\ s_\al s_\beta \al\ ,\ s_\al s_\beta s_\al \beta\ ,\ \ldots\ \bigr\}
\end{equation}
and reproduces the ordering of the corresponding Weyl reflections in~(\ref{Weyl}),
with $\gamma_0=-\beta$ and $\gamma_1=\al$. 
The real roots $\gamma_k$ are the integral points on the positive branch of the quadric~(\ref{quadric}),
while the negative branch contains the set~$\{-\gamma_k\}$.

Remembering $s_\gamma x=x-(x{\cdot}\gamma)\,\gamma$ we combine the reflections
\begin{equation}
\gamma_k\ \buildrel{s_\beta}\over{\longmapsto}\ -\gamma_{-k} \quad\und\quad
\gamma_k\ \buildrel{s_\al}\over{\longmapsto}\ -\gamma_{-k+2} \qquad\with\qquad
\gamma_k\ \buildrel{\al\leftrightarrow\beta}\over{\longmapsto}\ -\gamma_{1-k}
\end{equation}
and find for $\gamma_k=\xi_k\al+\eta_k\beta$ the recursion
\begin{equation}
\xi_{k+1}=m\,\xi_k-\eta_k \quad\und\quad \eta_{k+1}=\xi_k \qquad\for\qquad
\gamma_k = \xi_k \al + \eta_k \beta\ .
\end{equation}
This yields the three-term recursion relation
\begin{equation} \label{recursion}
\xi_{k+1} - m\,\xi_k + \xi_{k-1} = 0 \qquad\with\quad \xi_1=1 \und \xi_0=\eta_1=0\ .
\end{equation}
We note that the recursion can be iterated to the right as well as to the left, with
\begin{equation}
\xi_{-k} = -\xi_k \quad\und\quad \eta_{-k+1} = -\eta_{k+1}\ .
\end{equation}
One may check that
\begin{equation}
\gamma_k{\cdot}\gamma_{k'} \= 
2\,(\xi_k\xi_{k'}+\xi_{k-1}\xi_{k'-1}) - m\,(\xi_k\xi_{k'-1}+\xi_{k-1}\xi_{k'}) \=
\xi_{k-k'+1}-\xi_{k-k'-1}
\end{equation}
due to (\ref{recursion}), so that indeed $\gamma_k{\cdot}\gamma_k=\xi_1{-}\xi_{-1}=2$. 
The recursion can be solved explicitly,
\begin{equation} \label{xk1}
\xi_{k+1} \= \sum_{\ell=0}^{\lfloor k/2\rfloor} (-1)^\ell \binom{k{-}\ell}{\ell}\,m^{k-2\ell}
\qquad\for\quad k=0,1,2,3,\ldots
\end{equation}
giving
\begin{equation}
\xi_{k+1} \= 1\,,\,m\,,\,m^2{-}1\,,\,m^3{-}2m\,,\,m^4{-}3m{+}1\,,\,m^5{-}4m^3{+}3m\,,\,\ldots\ ,
\end{equation}
or via a generating function
\begin{equation}
F(z):=\sum_{k=0}^\infty \xi_{k+1}\,z^k \qquad\Rightarrow\qquad
F(z) \= \bigl(1-m\,z+z^2\bigr)^{-1}\ .
\end{equation}
The zeros of the characteristic polynomial provide a simple closed expression,
\begin{equation} \label{xk2}
z_\pm^2-m\,z_\pm+1=0 \qquad\Rightarrow\qquad
z_\pm = \tfrac12\bigl(m\pm\sqrt{m^2{-}4}\bigr) \qquad\Rightarrow\qquad
\xi_{k+1} \= \frac{1-z_\pm^{2k+2}}{z_\pm^k\ (1-z_\pm^2)}\ ,
\end{equation}
equally valid for both signs. 
Another useful parametrization of the real roots in $\Pi_m=\langle\al\beta\rangle$ is
\begin{equation}
\gamma_k \= \eta_k \bar\gamma + (\xi_k{-}\eta_k)\al  
\= \xi_{k-1} \bar\gamma + (\xi_k{-}\xi_{k-1})\al 
\qquad\with\quad \bar\gamma=\al{+}\beta\ ,
\end{equation}
which exhibits the symmetry axis $\bar\gamma$ of the quadric~(\ref{quadric}).

Equipped with these tools, we can further specify
\begin{equation}
Y \= \sum_{\{\Pi\}} Y^{\Pi} \= \sum_{m=0}^\infty \sum_{\{\Pi_m\}}  Y^{\Pi_m} 
\end{equation}
with, representing $\Pi_m=\langle\al\beta\rangle$, 
\begin{equation}
\begin{aligned}
Y^{\Pi_m}
&\= \tfrac12\,\al{\wedge}\beta {\sum_{k,k'\in\Z}}
\frac{(\xi_k\eta_{k'}-\xi_{k'}\eta_k)\ [ s_{\gamma_k},s_{\gamma_{k'}} ] }
{(\xi_k\,\al{\cdot}x + \eta_k\,\beta{\cdot}x)\,(\xi_{k'}\,\al{\cdot}x + \eta_{k'}\,\beta{\cdot}x)} \\
&\= \tfrac12\,\al{\wedge}\beta {\sum_{k,k'\in\Z}}
\frac{ \xi_{k'-k}\ \bigl\{ (s_\al s_\beta)^{k-k'} - (s_\beta s_\al)^{k-k'}\bigr\} }
{(\xi_k\,\al{\cdot}x + \xi_{k-1}\,\beta{\cdot}x)\,(\xi_{k'}\,\al{\cdot}x + \xi_{k'-1}\,\beta{\cdot}x)} \\
&\= -\al{\wedge}\beta\,\sum_{\ell=1}^\infty \sum_{k\in\Z} 
\frac{\xi_\ell\ \bigl\{ (s_\al s_\beta)^{\ell} - (s_\beta s_\al)^{\ell}\bigr\} }
{\bigl[ \xi_{k-1}\,\bar\gamma{\cdot}x+(\xi_k{-}\xi_{k-1})\,\al{\cdot}x \bigr]\
\bigl[ \xi_{k-\ell-1}\,\bar\gamma{\cdot}x+(\xi_{k-\ell}{-}\xi_{k-\ell-1})\,\al{\cdot}x \bigr] }\ ,
\end{aligned}
\end{equation}
where we used that $\xi_k\xi_{k'-1}{-}\xi_{k'}\xi_{k-1}$ 
does not change under a common shift of $k$ and~$k'$.

For $m{=}0$,
\begin{equation}
\xi_k \= \ldots\,,\,1\,,\,0\,,\,-1\,,\,0\,,\,1\,,\,0\,,\,-1\,,\,0\,,\,1\,,\,0\,,\,-1\,,\,\ldots\ ,
\end{equation}
and the situation is trivial since $\al{\perp}\beta$, thus $[s_\al,s_\beta]=0$ 
and no further roots are generated, so the subalgebra is $A_1\oplus A_1$.
For $m{=}1$, 
\begin{equation}
\xi_k \= \ldots\,,\,0\,,\,-1\,,\,-1\,,\,0\,,\,1\,,\,1\,,\,0\,,\,-1\,,\,-1\,,\,0\,,\,1\,,\,\ldots\ ,
\end{equation}
so the ellipse in~(\ref{quadric}) contains one additional root $\al{+}\beta$ (and all negatives),
making up an $A_2$ subalgebra. Its contribution to~$Y$ is proportional to
\begin{equation}
\frac{1}{\al{\cdot}x\ \beta{\cdot}x} + 
\frac{1}{\beta{\cdot}x\ \gamma{\cdot}x} + 
\frac{1}{\gamma{\cdot}x\ \al{\cdot}x} \= 
\frac{ (\al{+}\beta{+}\gamma){\cdot}x}{\al{\cdot}x\ \beta{\cdot}x\ \gamma{\cdot}x} \= 0
\qquad\for\quad \gamma=-(\al{+}\beta)\ .
\end{equation}
Therefore, $Y^{\Pi_m}$ vanishes for $m=0$ and~$1$.
The corresponding finite root systems are shown in Figure~12.
\begin{figure}[h!tbp]
\centering
\includegraphics[width = 0.35\paperwidth]{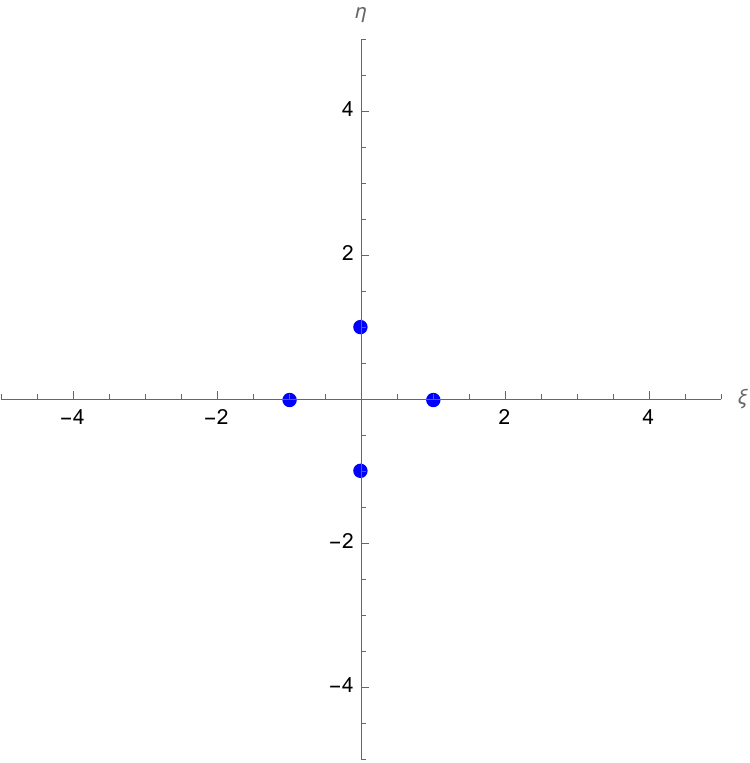} \qquad
\includegraphics[width = 0.35\paperwidth]{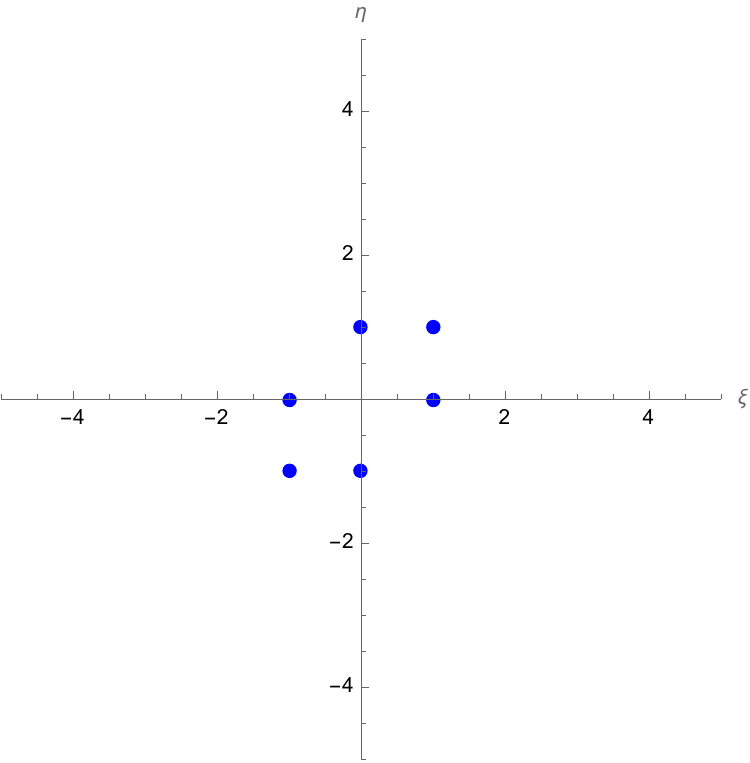}
\caption{Real root system $\{\gamma=\xi\al+\eta\beta\}$ for ``elliptic planes'' 
from $|\al{\cdot}\beta|=0$ (left) or $1$ (right)}
\end{figure}
A more interesting case occurs for $m{=}2$. 
Here, the real roots lie on two straight lines (see Figure~13 left), 
\begin{equation}
\xi_k \= k \qquad\Rightarrow\qquad
\pm\gamma_k \= k\,\al + (k{-}1)\beta \= (k{-}1)\bar\gamma + \al
\end{equation}
where $\bar\gamma=\al{+}\beta$ happens to be null and orthogonal to $\al$ and~$\beta$.
This set of roots creates the affine extension 
$\widehat{sl}_2\equiv\widehat{A}_1\equiv A_1^{(1)}$ of $sl_2$.
Obviously, any pair of roots in this set has a scalar product of $2$ or $-2$.
Such an $\widehat{sl}_2$ subsystem is generated by any non-orthogonal pair of real roots
from levels $\ell{=}{\pm}1$ and $|\ell|{\le}1$.
Its contribution to $Y$ evaluates to
\begin{equation}
\begin{aligned}
Y^{\Pi_2} &\=
-\al{\wedge}\beta\,\sum_{\ell=1}^\infty \ell\,\bigl\{(s_\al s_\beta)^\ell-(s_\beta s_\al)^\ell\bigr\}
\sum_{k\in\Z} \bigl[ k\,\bar\gamma{\cdot}x+\al{\cdot}x \bigr]^{-1}  
\bigl[ (k{-}\ell)\,\bar\gamma{\cdot}x+\al{\cdot}x \bigr]^{-1} \\
&\= \al{\wedge}\beta\,\sum_{\ell=1}^\infty \ell\,\bigl\{(s_\al s_\beta)^\ell-(s_\beta s_\al)^\ell\bigr\}
\,\frac{\pi}{\ell\,(\bar\gamma{\cdot}x)^2}\,
\Bigl\{ \cot\bigl(\pi[\ell{-}\tfrac{\al{\cdot}x}{\bar\gamma{\cdot}x}]\bigr) 
+ \cot\bigl(\pi[\tfrac{\al{\cdot}x}{\bar\gamma{\cdot}x}]\bigr)\Bigr\} \= 0
\end{aligned}
\end{equation}
due to $\cot(x)+\cot(\ell\pi{-}x)=0$. 
Hence, also the affine subalgebras do not obstruct a commutativity of the linear Dunkl operators.

As soon as we go beyond level one, real root pairs with $m{>}2$ show up,
and the associated quadric~(\ref{quadric}) is a hyperbola (see Figure~13 right for $m{=}3$). 
Let us inspect the simplest such case, $m{=}3$,
where the coefficient sequence happens to be the even half of the {\it Fibonacci sequence\/}
($k=1,2,3,\ldots$),
\begin{equation}
\xi_{k+1}=3\,\xi_k-\xi_{k-1} \with \xi_1{=}1 \quad\&\quad \xi_0{=}0 \quad\Rightarrow\quad
\xi_k = 1, 3, 8, 21, 55, 144, 377, 987, \ldots = f_{2k}\ ,
\end{equation}
where $f_{n+1}=f_n+f_{n-1}$ with $f_0=0$~\cite{Feingold,Penta}.
The root scalar products take the values $\pm2,\pm3,\pm7,\pm18,\pm47$ etc..
\begin{figure}[h!tbp]
\centering
\includegraphics[width = 0.35\paperwidth]{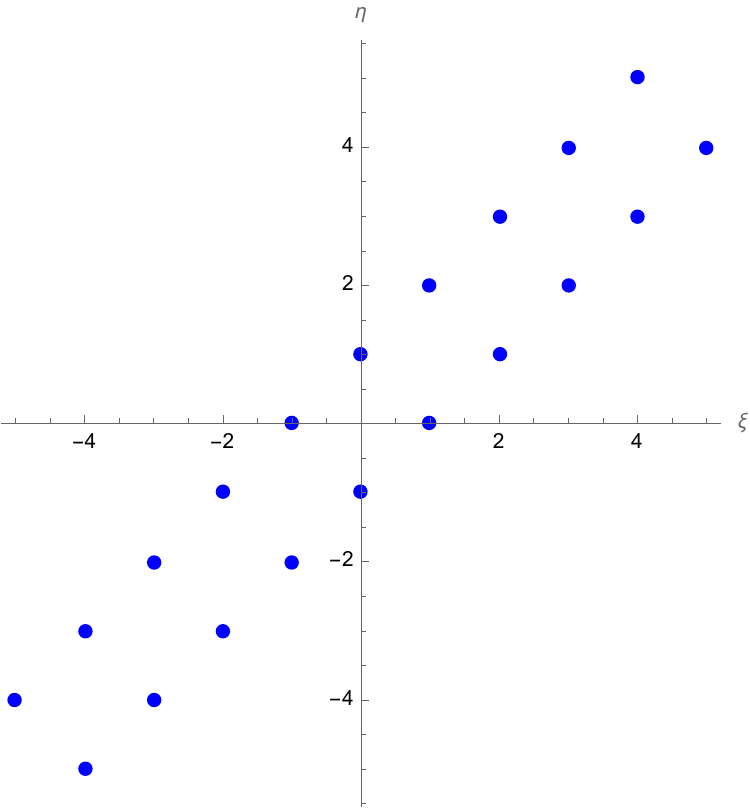} \qquad
\includegraphics[width = 0.35\paperwidth]{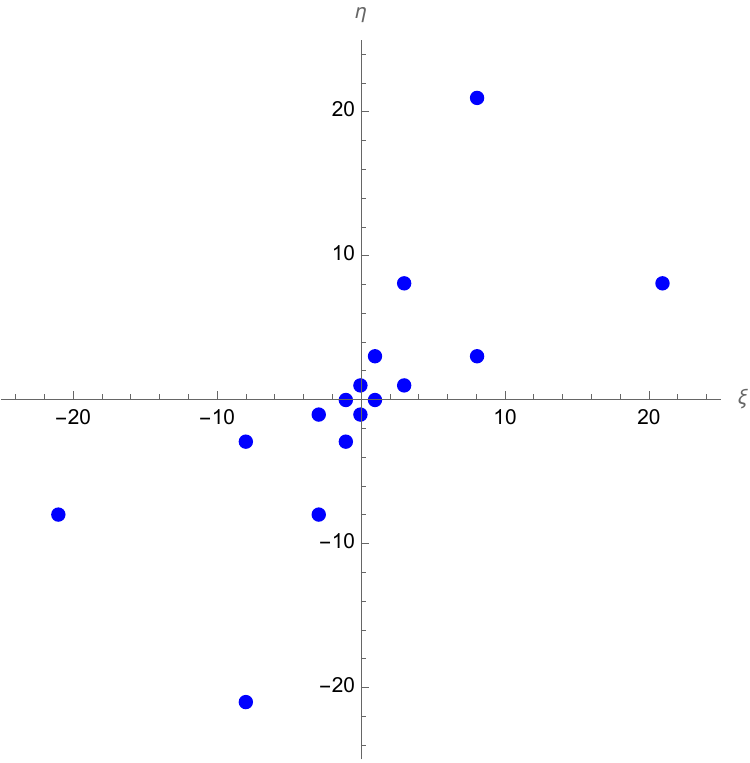}
\caption{Real roots for a ``parabolic plane'' ($|\al{\cdot}\beta|{=}2$, left) 
and for a ``hyperbolic plane'' ($|\al{\cdot}\beta|{=}3$, right)}
\end{figure}
In this case, the contribution to the two-form~$Y$ becomes
\begin{equation}
\begin{aligned}
Y^{\Pi_3} &\=
-\al{\wedge}\beta\,\sum_{\ell=1}^\infty \sum_{k\in\Z} 
\frac{f_{2\ell}\ \bigl\{ (s_\al s_\beta)^{\ell} - (s_\beta s_\al)^{\ell}\bigr\} }
{\bigl[ f_{2k}\,\bar\gamma{\cdot}x+f_{2k+1}\,\al{\cdot}x \bigr]\
\bigl[ f_{2k-2\ell}\,\bar\gamma{\cdot}x+f_{2k-2\ell+1}\,\al{\cdot}x \bigr] }\ ,
\end{aligned}
\end{equation}
with the understanding that $f_{-n}=(-1)^{n+1}f_n$ extends the Fibonacci sequence to the left.
As numerical checks show, the individual sums over~$k\in\Z$ do not vanish, 
nor does the total expression. 
Turning off one of the two Weyl orbits in the root-sum for the Dunkl operator~(\ref{Dunkl})
does not help since odd values of~$m$ require both $\alpha$ and $\beta$ to lie in~${\cal R}_+$,
and thus only this orbit contributes here.
We are forced to conclude that $Y\neq 0$ for our model,
so its linear Dunkl operators~${\cal D}_\mu$ do not commute to any simple expression, 
and so we cannot construct higher conserved charges in this way.
Likewise, $Y_\mu^{ \ \mu}$ does not vanish either, and thus $\textrm{res}\sum{\cal L}^2$
in (\ref{L2hyp}) does not reproduce the Hamiltonian~$H_\Omega$.
This points to a lack of integrability, but falls short of disproving it.

\newpage
\section{Conclusions}
\noindent
Spherical angular Calogero models are obtained by reducing a rational Calogero model in $\R^n$ 
to the sphere~$S^{n-1}$. Analogously, we have defined a hyperbolic angular model by reducing
a Calogero Hamiltonian in $\R^{1,n-1}$ to the (future) hyperboloid~$H^{n-1}$.
The main difference to the conventional angular model is the non-compactness of hyperbolic space
and the replacement of a finite spherical Coxeter reflection group by an infinite hyperbolic one.
As a consequence, the Calogero-type potential of the model is an infinite sum over 
all hyperbolic reflections and not easily obtained in a closed form. 
However, it is a real automorphic function of an associated hyperbolic Kac--Moody algebra.

We have worked out the details for the rank-3 case of $AE_3$ leading to a PGL(2,$\Z$)
invariant quantum mechanical model on the Poincar\'e disk or the complex upper half plane.
In this case, the potential can be reformulated as a Poincar\'e series, which converges outside
the mirror lines of PGL(2,$\Z$). We then asked whether the integrability of the spherical angular 
models extends to the hyperbolic ones. To this end, we introduced the Dunkl operators
for the $AE_3$ algebra, {\bf on $\R^{1,2}$ and on $H^2$,} 
and computed their commutators. It turned out that the presence of
hyperbolic rank-2 subalgebras in $AE_3$ prevents a simple result (like zero).
This may be an obstacle to integrability, but is not enough to rule it out.

We comment that the energy spectrum of the $AE_3$ hyperbolic Calogero model is a deformation
of the discrete parity-odd part of the spectrum of the hyperbolic Laplacian on square-integrable
automorphic functions, because the singular lines of the potential impose Dirichlet boundary
conditions on the boundary of the fundamental domain. It remains to be seen whether
the spectral flow with the coupling~$g$ is isospectral under integral increments of~$g$.

Since the Weyl-alcove walls of certain hyperbolic Kac--Moody algebras are the cushions
of the billard dynamics in the BKL approach~\cite{BKL1,BKL2}, 
the small-$g$ limit of our hyperbolic Kac--Moody Calogero system provides a model
for cosmological billards~\cite{DHN}. The chaotic dynamics of such billards seems to be consistent 
with a formal integrability of the corresponding hyperbolic Toda-like theories~\cite{DHN}. 
Furthermore, an alternative description of BKL dynamics leads instead to
Euler--Calogero--Sutherland potentials of the $\sinh^{-2}$ type, which also produce 
sharp walls in the BKL limit~\cite{Ryan,KhvMla}.
This nurtures the hope that also our Calogero-type potentials retain a kind of integrability.
Finally, the well-known quantum chaotic behavior of hyperbolic billards~\cite{Bogo} 
may be ``tamed'' by turning on our Kac--Moody Calogero potential, 
since in the large-$g$ limit the wave function will get pinned near the bottom of the potential.
We hope that this opens a door to interesting further studies in this field.

\newpage
\appendix
\section{Appendix (with Don Zagier): Computing the potential}
\noindent
In this appendix we investigate the numerical evaluation of the potential function~$U(z)$.
We have to sum over infinitely many triples $(\ell,m,n)$ or $(p,q,r)$ subject to a diophantine equation,
see~(\ref{mirrorpqr})--(\ref{mirrorlmn}),
\begin{equation} \label{Upqr}
U(z) =\!\!\!\sum_{\genfrac{}{}{0pt}{2}{p,q,r}{p^2+qr=1}} \!\!\!\frac{y^2}{[r(x^2{+}y^2)-2p\,x-q]^2}
\= \Bigl(\sum_{r=0}+\,2\!\sum_{r>0}\Bigr) 
\!\!\!\sum_{\genfrac{}{}{0pt}{2}{p,q}{p^2+qr=1}} \!\!\!\frac{y^2}{[r(x^2{+}y^2)-2p\,x-q]^2}
\,=:\, U_0(z)+2\,U_>(z)\,,
\end{equation}
where we used that $(p,q,r)$ and $(-p,-q,-r)$ contribute equally.
The $r{=}0$ part is easily summed up since then $p=\pm1$, which yields
\begin{equation}
U_0(z)\ :=\ 2 \sum_{q\in\Z} \frac{y^2}{[2\,x+q]^2} \= \frac{2\pi^2y^2}{\sin^2(2\pi x)}\ .
\end{equation}

For $U_>$ we do not know how to compute the sum in closed form.
The decomposition of the root space detailed in Section~2 suggests to slice the space of triples
according to fixed values of
\begin{equation}
\ell = r-p-q \qquad\Rightarrow\qquad q = r-p-\ell\ ,
\end{equation}
in which case the discussion there shows that for each fixed value of $\ell$ we are
left with only a sum over finitely many pairs $(p,r)$.  Specifically,
if we rewrite the diophantine equation $p^2+qr=1$ as
\begin{equation}
3\,r^2 + (r-2p)^2 \= 4(\ell\,r+1)  \quad  \le\ 4\,r\,(\ell{+}1) \ ,
\end{equation}
then we see that for $r>0$ we must have $\ell\ge0$ and also $r\le4(\ell+1)/3$,
reducing the sum over~$r$ to a finite one.
Then only those $r$ for which $4\ell\,r+4-3r^2$ is a square give a contribution.
Thus we obtain
\begin{equation} \label{Ul}
U_>(z) \= \sum_{\ell=0}^\infty \sum_{r=1}^{\lfloor4(l+1)/3\rfloor} \!\!\sum_{s^2=4\ell r+4-3r^2} 
\frac{y^2}{[r(x^2{+}y^2)-(r{-}s)\,x-\frac{r+s}2{+}\ell]^2}\ ,
\end{equation}
where the inner sum is almost always empty and never has more than two terms.

The expression~(\ref{Ul}) converges rather slowly. But one can do better,
by going back to~\eqref{Upqr} and reducing the inner sum to a finite one in the following way.
For each fixed $r{>}0$ we denote this inner sum by $U_r(z)$ and rewrite it as
\begin{equation} \label{Ur}
U_r(z) \= \frac{y^2}{r^2}
\!\!\!\sum_{\genfrac{}{}{0pt}{2}{p\,\in\,\Z}{p^2\equiv1\,(\textrm{mod}\,r)}}\!\!
\frac1{\bigl[ (x-\tfrac{p}{r})^2 + y^2 - \tfrac{1}{r^2}  \bigr]^2}
\= \frac{y^2}{r^2}
\!\!\!\sum_{\genfrac{}{}{0pt}{2}{p\,(\textrm{mod}\,r)}{p^2\equiv1\,(\textrm{mod}\,r)}}\!\!\!
S\bigl(x-\tfrac{p}{r},\sqrt{\vphantom{|}\smash{ y^2-\tfrac{1}{r^2}}} \bigr)
\end{equation}
where the function~$S$ is defined on $\C{\times}\C$ \
by $S(x,a):=\sum_{n\,\in\,\Z} \bigl[(x-n)^2+a^2\bigr]^{-2}$,
which obviously depends only on $x$~(mod~1). Using a partial fraction
decomposition of the summand together with Euler's formul\ae\ for $\sum_{n\in\Z}1/(x{+}n)$
and $\sum_{n\in\Z}1/(x{+}n)^2$, we find the closed formula
\begin{equation} \label{S}
S(x,a) \= \frac{\pi}{2\,a^3}\,\frac{\sinh(2\pi\,a)}{\cosh(2\pi\,a)-\cos(2\pi\,x)\vphantom{\big|^2}}
\ +\ \frac{\pi^2}{a^2}\,\frac{\cosh(2\pi\,a)\,\cos(2\pi\,x)-1}{\bigl(\cosh(2\pi\,a)-\cos(2\pi\,x)\bigr)^2}\ .
\end{equation}
Inserting this into~(\ref{Ur}) then expresses each $U_r(z)$ as a finite sum of elementary functions,
and formula~(\ref{Upqr}) takes on the more explicit form
\begin{equation} \label{Uresult}
U(z) \= \frac{2\pi^2\,y^2}{\sin^2(2\pi x)} \ +\ 2\,\sum_{r=1}^\infty \ \frac{y^2}{r^2}
\!\!\!\sum_{\genfrac{}{}{0pt}{2}{p\,(\textrm{mod}\,r)}{p^2\equiv1\,(\textrm{mod}\,r)}}\!\!\!
S\bigl(x-\tfrac{p}{r},\sqrt{\vphantom{|}\smash{ y^2-\tfrac{1}{r^2}}} \bigr)
\end{equation}
in which we now simply {\it define\/} $S(x,a)$ by the trigonometric formula~\eqref{S}.

Formula~\eqref{Uresult} is both simpler and more rapidly convergent than the original
expression~\eqref{Upqr}, since the internal infinite sums have been evaluated explicitly, and it also
converges more rapidly than the ``slicing by~$\ell\,$" formula~(\ref{Ul}). As a demonstration, we list
in Table~2 ten-digit values of the partial sums $U(z,R)$ defined by truncating~\eqref{Uresult}
at $r{=}R$ for a typical point $z_1=0.1+0.7\,\im$ and values of~$R$ going up to one million. We have also 
included the values $U(z_2,R)$ at the modular image $z_2=-1/z_1=-0.2+1.4\,\im$, 
both as a test of the modularity of~$U(z)$
and as a confirmation of the accuracy of the computation.
\begin{table}[h!]
{\small
\begin{center}
\begin{tabular}{|c|ccccccc|}
\hline
$R$ & $10\,000$ & $20\,000$ & $50\,000$ & $100\,000$ & $200\,000$ & $500\,000$ & $1\,000\,000$ \\
\hline
$\rs U(z_1,R)\rs\phantom{\Big|}$ & 
52.24167922 \rs&\rs 
52.24327256 \rs&\rs 
52.24429208 \rs&\rs 
52.24465475 \rs&\rs 
52.24484553 \rs&\rs 
52.24496635 \rs&\rs 
52.24500890 \rs\\ 
$\rs U(z_2,R)\rs$ & 
52.24339662 \rs&\rs 
52.24417862 \rs&\rs  
52.24467954 \rs&\rs  
52.24485793 \rs&\rs  
52.24495185 \rs&\rs 
52.24501138 \rs&\rs  
52.24503236 \rs\\
\hline
\end{tabular}
\caption{Values of truncated sums for the potential, at $z_1=0.1+0.7\,\im$ and at $z_2=-1/z_1$}
\end{center}
}
\vspace{-5mm}
\end{table}
\noindent 
It takes {\tt PARI} about 17 minutes for $R{=}10^5$ and about 28 hours for $R{=}10^6$ 
on a standard workstation to compute the values for each point~$z_i$  given in this table. 
The output suggests that the final numbers are correct to about 7 significant digits.

The very erratic dependence of the numbers $U_r(z)$ on $r$,
due to the sum over square-roots of~$1$ modulo~$r$ in (\ref{Ur}),
prohibits further analytic simplification. For the same reason, 
the infinite sum for $U(z)$, although convergent, is not very tractable numerically.
However, the convergence of the partial sums $U(z,R)$ for $R\to\infty$
can be accelerated by adding a suitable correction term. The following
heuristic argument suggests what this correction term should be.
If the inner sum in~\eqref{Uresult} were over all values of~$p$ (mod~$r$) then,
since the value of  $\sqrt{\vphantom{|}\smash{ y^2-\tfrac{1}{r^2}}}\approx y$
is close to~$y$ for $r$ large, this inner sum for
large~$r$ would simply be $r$ times a Riemann sum for the integral 
$\int_{\R/\Z}S(x,y)\,\diff x=\int_\R(x^2{+}y^2)^{-2}\diff y=\pi/2y^3$ and hence could be 
approximated by~$r\pi/2y^3$.  The actual inner sum is only over $N(r)$ rather
than $r$ values of $p$~(mod~$r$), where $N(r)$ denotes the number of square-roots
of~$1$ modulo~$r$. Hence, if these square-roots are more or less uniformly distributed
on the interval $[1,r]$ on the average, which is a reasonable heuristic assumption,
then the value of the inner sum should be roughly $N(r)\,\pi/y^3$ on average.
Therefore, the contribution of the terms in~\eqref{Uresult} with $r{>}R$ (the ``tail'') 
should be approximately $\pi/y$ times $\sum_{r>R}N(r)/r^2$ for $R$ large.  The
value of the arithmetic function $N(r)$ fluctuates a lot, but its average behavior is
quite regular, and one can give the asymptotic value of the sum $\sum_{r>R}N(r)/r^2$
without difficulty.  Specifically, from the Chinese remainder theorem we find that $N(r)$
is multiplicative, meaning that $N\bigl(\prod p_i^{\nu_i}\bigr)=\prod N(p_i^{\nu_i})$,
and $N(p^\nu)$ in turn is easily evaluated as~2 for $p$ an odd prime and $\nu{\ge}1$
(the only two square-roots of~1 in this case being $\pm 1$~(mod~$p^r$)), and as 1 or 2 or~4
for $p{=}2$ and $\nu{=}1$, $2$, or ${\ge}3$, respectively
(the only square-roots of~1 in the latter case being $\pm1$ and $\pm 1{+}2^{r-1}$~(mod~$2^r$)).
This gives
\begin{equation}
\begin{aligned}
{\cal N}(s)\ &:=\
\sum_{r=1}^\infty \frac{N(r)}{r^s} \= 
\Bigl(1 + \frac{1}{2^s} + \frac{2}{4^s} + \frac{4}{8^s} + \frac{4}{16^s} + \ldots\Bigr)\,
 \prod_{p>2} \Bigl(1 + \frac{2}{p^s} + \frac{2}{p^{2s}} + \frac{2}{p^{3s}} + \ldots\Bigr) \\
&\= \frac{1\,+\,2^{-2s}+2^{1-3s}}{1\,-\,2^{-s}}\,\prod_{p>2} \frac{1+p^{-s}}{1-p^{-s}}
\= \bigl(1\,-\,2^{-s}+2^{1-2s}\bigr)\,\frac{\z(s)^2}{\z(2s)}\ ,
\end{aligned}
\end{equation}
where $\z(s)$ denotes the Riemann zeta function. In particular, $\mathcal N(s)$ has a double pole
at $s{=}1$ with principal part $\frac{\z(2)^{-1}}{(s-1)^2}\+\text O\bigl(\frac1{s-1}\bigr)$, so $N(r)$ 
behaves ``on the average" like $\z(2)^{-1}\log r$, and $\sum\limits_{r>R}\frac{N(r)}{r^2}$ is asymptotically
equal to $\z(2)^{-1}\,\frac{\log R}R$. This suggests that we can improve the convergence of $\sum U_r(z)$ 
by replacing the partial sums $U(z,R)$ by
\begin{equation}\label{defU0}
U^{(0)}(z,R) \= U(z,R)\ +\ \frac6{\pi y}\,\frac{\log R}{R}\ .
\end{equation}
This is indeed the case, as we see from Table~3, in which we have tabulated the
corrected partial sums with the same parameters as before.
\begin{table}[htbp]
{\small
\begin{center}
\begin{tabular}{|c|ccccccc|}
\hline
$R$ & $10\,000$ & $20\,000$ & $50\,000$ & $100\,000$ & $200\,000$ & $500\,000$ & $1\,000\,000$ \rs\\
\hline
$\rs U^{(0)}(z_1,R)\rs\phantom{\Big|}$ &
52.24419214 \rs&\rs 
52.24462358 \rs&\rs 
52.24488249 \rs&\rs 
52.24496886 \rs&\rs 
52.24501205 \rs&\rs 
52.24503796 \rs&\rs 
52.24504659 \rs\\
$\rs U^{(0)}(z_2,R)\rs$ & 
52.24465308 \rs&\rs  
52.24485413 \rs&\rs 
52.24497474 \rs&\rs 
52.24501499 \rs&\rs 
52.24503511 \rs&\rs 
52.24504718 \rs&\rs 
52.24505121 \\
\hline
\end{tabular}
\caption{Values of firstly improved truncated sums for the potential, at $z_1=0.1+0.7\,\im$ and $z_2=-1/z_1$}
\end{center}
}
\vspace{-3mm}
\end{table}
We can improve these values further by adding an appropriate term $C/R$ to~\eqref{defU0}, where $C$ is 
a constant depending on~$z$ but not on~$R$.  For this, we must first give a more precise estimate of 
$\sum_{r>R} N(r)/r^2$ for $R$~large.  
The function $\z(s)$ is holomorphic except  for a simple pole at~$s{=}1$, with 
$\z(1{+}\ve)=\ve^{-1}+\gamma+\textrm{O}(\ve)$ as~$\ve\to0$, where $\gamma$ is Euler's constant. 
It has no zeros in the half-plane $\Re(s)>1$ (or even $\Re(s)>\frac12$ if we assume the Riemann Hypothesis), 
so $\mathcal N(s)$ has the same poles as $-c\;\z'(s)+c_1\,\z(s)$ in the half-plane $\Re(s)>\frac12$ 
(resp.~$>\frac14$ on RH), where
\begin{equation}
c\=\frac1{\z(2)}\=\frac6{\pi^2} \= 0.6079271\dots \quad\text{and}\quad
c_1\= c\,\Bigl(2\gamma\,-\frac12\,\log2 \,-\,2\,\frac{\z'(2)}{\z(2)}\Bigr) \= 1.184108\cdots\ .
\end{equation}
This implies that $N(r)$ behaves on the average like $c\,\log\,r+ c_1$, 
and that we have the asymptotic estimate
\begin{equation} \label{Ntail}
\sum_{r=R+1}^\infty \frac{N(r)}{r^2} \ \sim\ \frac{c\,\log R +c+c_1}{R}\ ,
\end{equation}
with an error of the order of $R^{-3/2+\ve}$ unconditionally or
$R^{-7/4+\ve}$ if we assume the Riemann Hypothesis.
This suggests that we can further improve the convergence of $\sum U_r(z)$ by replacing~$U^{(0)}(z,R)$ 
of~\eqref{defU0} with
\begin{equation}\label{defU1}
U^{(1)}(z,R) \= U^{(0)}(z,R) \ +\ \frac{\pi}{y}\,\frac{c+c_1}{R}\ ,
\end{equation}
and this is indeed confirmed by the values shown in Table~4.
\begin{table}[htbp]
{\small
\begin{center}
\begin{tabular}{|c|ccccccc|}
\hline
$R$ & $10\,000$ & $20\,000$ & $50\,000$ & $100\,000$ & $200\,000$ & $500\,000$ & $1\,000\,000$ \rs\\
\hline
$\rs U^{(1)}(z_1,R)\rs\phantom{\Big|}$ & 
52.24499640 \rs&\rs 
52.24502571 \rs&\rs 
52.24504334 \rs&\rs 
52.24504929 \rs&\rs 
52.24505226 \rs&\rs 
52.24505404 \rs&\rs 
52.24505464 \rs\\
$\rs U^{(1)}(z_2,R)\rs$ & 
52.24505521 \rs&\rs 
52.24505520 \rs&\rs 
52.24505517 \rs&\rs 
52.24505520 \rs&\rs 
52.24505522 \rs&\rs 
52.24505522 \rs&\rs 
52.24505523 \\
\hline
\end{tabular}
\caption{Values of secondly improved truncated sums for the potential, at $z_1=0.1+0.7\,\im$ and $z_2=-1/z_1$}
\end{center}
}
\vspace{-3mm}
\end{table}
However, although these values are better, we see clearly from the graphs shown in Figure~14  
that even $U^{(1)}(z_1,R)$, though much more nearly constant than $U^{(0)}(z_1,R)$, 
is still off by a linear term in~$1/R$.
\begin{figure}[htbp]
\begin{minipage}{0.31 \paperwidth}
   \includegraphics[scale=0.31]{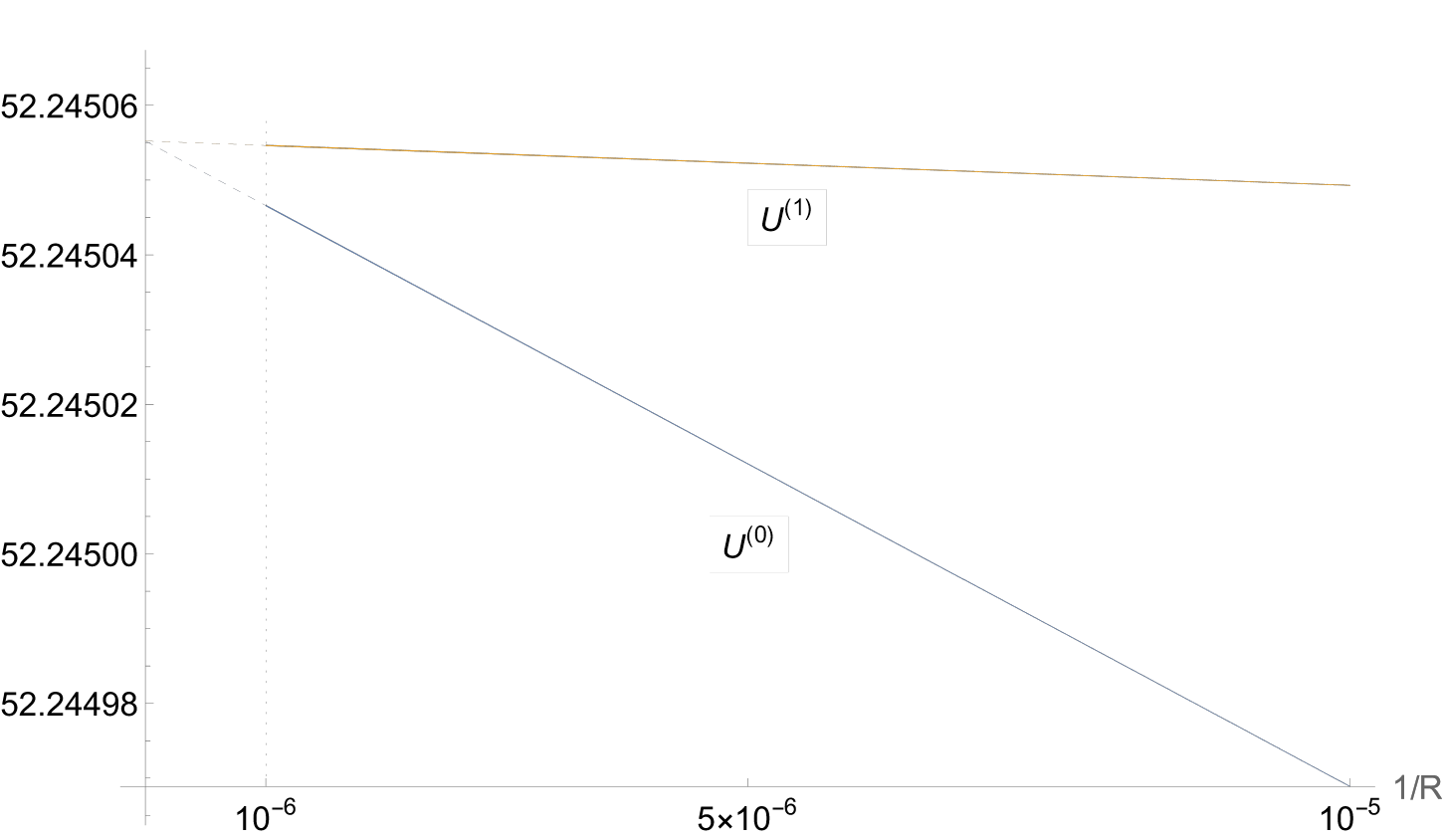} 
    \caption{Plot of the improved truncated sums $U^{(0)}(z_1,1/R)$ and $U^{(1)}(z_1,1/R)$}
\end{minipage}
\qquad\qquad\qquad
\begin{minipage}{0.31\paperwidth}
   \includegraphics[scale=0.31]{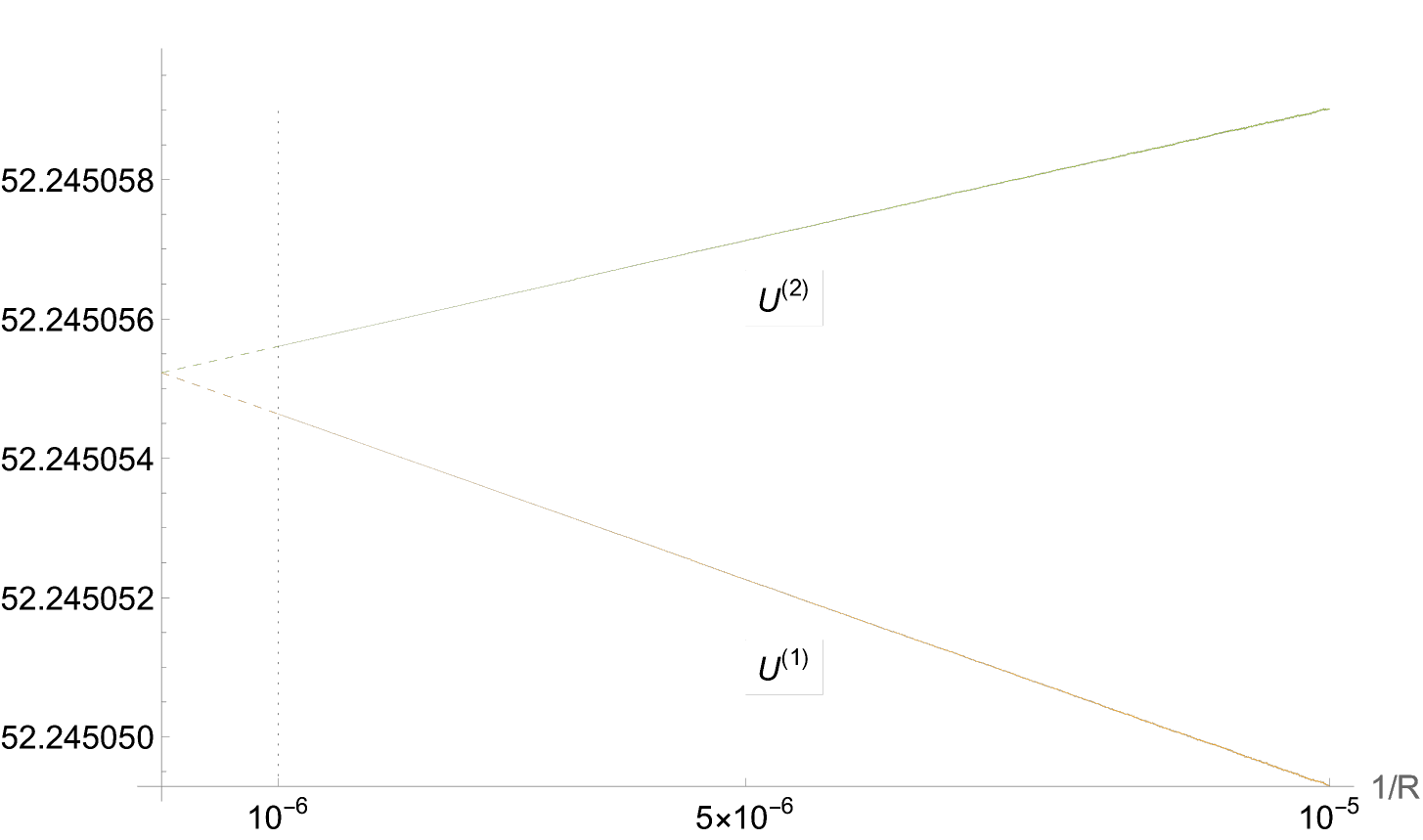}
    \caption{Plot of the improved truncated sums $U^{(1)}(z_1,1/R)$ and $U^{(2)}(z_1,1/R)$}
\end{minipage}
\end{figure}

To understand the reason for this, we must refine the heuristic argument given above.  
We begin by noting that the values of $\frac{p}{r}$ for $p^2\equiv1$~(mod~$r$) are in fact {\it not\/} 
completely uniformly distributed modulo~1, even on the average, 
because the two values corresponding to~$p\equiv\pm1$~(mod~$r$) are always very near to~0.  
For the remaining $N(r){-}2$ values the heuristic assumption of equidistribution at first sight 
still seems plausible, in which case the corresponding contribution to each term $U_r(z)$ with $r$ large 
would have the same average behavior as $\bigl(N(r){-}2\bigr)\frac\pi{r^2y}$, 
but the two values of $p/r$ near~0 give a contribution to~$U_r(z)$  of approximately $2y^2S(x,y)/r^2$ each.  
This suggests the improved correction
\begin{equation}\label{defU2}
U^{(2)}(z,R) \= U^{(1)}(z,R) \ +\ \frac{4y^2 S^*(x,y)}{R} \qquad\text{with}\qquad S^*(x,y):=S(x,y)-\frac\pi{2y^3}
\end{equation}
instead of~(\ref{defU1}), and this indeed does give a further improvement of the convergence, 
as one sees in Table~5 and also in the graph in Figure~15.  
But this same graph makes it clear that we still do not have the right linear correction 
term. The reason for this is more subtle and is the most interesting part of the discussion.
\begin{table}[htbp]
{\small
\begin{center}
\begin{tabular}{|c|ccccccc|}
\hline
$R$ & $10\,000$ & $20\,000$ & $50\,000$ & $100\,000$ & $200\,000$ & $500\,000$ & $1\,000\,000$ \\
\hline
$\rs U^{(2)}(z_1,R)\rs\phantom{\Big|}$ & 
52.24509364 \rs&\rs 
52.24507433 \rs&\rs  
52.24506279 \rs&\rs 
52.24505901 \rs&\rs 
52.24505712 \rs&\rs 
52.24505599 \rs&\rs 
52.24505561 \rs\\
$\rs U^{(2)}(z_2,R)\rs$ & 
52.24505562 \rs&\rs 
52.24505540 \rs&\rs 
52.24505525 \rs&\rs 
52.24505524 \rs&\rs 
52.24505524 \rs&\rs 
52.24505523 \rs&\rs 
52.24505523 \rs\\
\hline
\end{tabular}
\caption{Values of thirdly improved truncated sums for the potential, at $z_1=0.1+0.7\,\im$ and $z_2=-1/z_1$}
\end{center}
}
\vspace{-5mm}
\end{table}

This final heuristic argument depends on the observation 
that not only are the two obvious square-roots~$\pm1$ of 1~(mod~$r$) not randomly distributed modulo~$r$, 
but that there are infinitely many other non-randomly distributed square-roots, 
each occurring for a set of integers~$r$ of positive asymptotic density.
For instance, if $r\!\equiv10\!\pmod{25}$, which happens for $4\%$ of all integers, 
then the two further numbers $\pm{(r/5\,-\,1)}$ are also square-roots of 1~(mod~$r$), 
because $ r/5\,-\,1$ is congruent to $-1$ modulo~$r/5$ and to~$+1$ modulo~5.  
For these two values of~$p$, one has $\frac pr\equiv\pm\frac15+\text O\bigl(\frac1r\bigr)$,
which are indeed not randomly distributed.  
More generally, for any rational number $\al=N/D$ with $N$ and~$D$ coprime
and~$D>0$, we consider integers $r{>}0$ of the form $D\,n$ with $n\equiv-2N^{-1}\!\pmod D$.   
Then the number $p=Nn+1$ is congruent to~1 modulo~$n$ and to~$-1$ modulo~$D$, 
so that $p^2\equiv1\!\!\pmod r$, while $\frac pr=\frac ND+\frac1r$ is extremely close to~$\al$ if $r$ is large.   
For fixed~$\al$, the set of integers~$r$ of this form constitutes a single congruence class modulo~$D^2$ 
and hence has asymptotic density~$1/D^2$. Hence, approximating $x-\frac pr$ by~$x-\al$
and $\sqrt{\vphantom{|}\smash{ y^2-\tfrac{1}{r^2}}}$ by~$y$ in~\eqref{Ur} ,
we see that the total contribution of these terms to the ``tail" $r>R$ 
in~\eqref{Ur} is approximately $\frac{2\,y^2}{D^2 R}\,S(x-\al,y)$.
To get the final answer, we must sum this over all rational numbers~$\al\!\pmod1$, 
{\rm i.e.}, over all denominators $D>0$ and all numerators $N\!\pmod D$ prime to~$N$.

The easiest way to do this is to use the Fourier expansion of the periodic function $S(x,a)$ which is  given by 
\begin{equation}\label{Four} 
S(x,a) \= \frac\pi{a^3}\,\Bigl(\frac12 \+ \sum_{n=1}^\infty \bigl(1+2\pi na\bigr)\,\ep^{-2\pi na}\,\cos(2\pi nx)\Bigr)\ .
\end{equation}
(There are two ways to see this: 
either one writes the Fourier expansion of $S(x,a)$ as $\sum_{n\in\Z}S_n(a)\,\ep^{2\pi inx}$ 
with $S_n(a)=\int_{\R/\Z} S(x,a)\,\ep^{-2\pi nx}\,\diff x=\int_\R \frac{\ep^{-2\pi nx}}{(x^2+a^2)^2}\,\diff x$ 
and computes the integral by the Cauchy residue theorem, 
or else one simply evaluates the expression on the right-hand side of~\eqref{Four} in closed form 
using the formul\ae\ for the sum of a geometric series or its derivative, 
obtaining precisely the right-hand side of~\eqref{S}.)
The constant term $\pi/2a^3$, in this expansion, with $a$ 
replaced by $\sqrt{\vphantom{|}\smash{ y^2-\tfrac{1}{r^2}}}\approx y$ and inserted into~\eqref{Uresult}, 
gives exactly the approximation $\frac\pi y\,\sum_{r>R}\frac{N(r)}{r^2}$ to $U(z)-U(z,R)$
that we used in our initial heuristic argument and that led via~\eqref{Ntail}
to the function $U^{(1)}(z,R)$ defined in~\eqref{defU1}.  
The further correction term coming from all rational numbers $\al\!\pmod1$ as explained above therefore
has the form $C(z)/R$, with $C(z)$ given by
\begin{equation}
C(x{+}\im y) \= 2y^2\,\sum_{D=1}^\infty \frac1{D^2}\!\!
\sum_{\genfrac{}{}{0pt}{2}{N\,\text{(mod~$D$)}}{(N,D)=1}} \!\! S^*\bigl(x{-}\tfrac ND,y\bigr)
\end{equation}
with  $S^*(x,a)$ as in~\eqref{defU2}.  
Replacing $S^*(x,a)$ by its expression as a sum of exponentially small terms
given  in~\eqref{Four} and interchanging the order of summation, we find
\begin{equation}
C(x{+}\im y) \,=\,\frac{2\pi}y \sum_{n=1}^\infty \ve(n)\,(1{+}2\pi ny)\,\ep^{-2\pi ny}\,\cos(2\pi nx)
\qquad\text{with}\qquad
\ve(n):= \sum_{D=1}^\infty \frac1{D^2}\!\!\sum_{\genfrac{}{}{0pt}{2}{N\,\text{(mod~$D$)}}{(N,D)=1}}\!\!
\ep^{2\pi\im nN/D}\ .
\end{equation}
The sum over $N\!\pmod D$ is the well-known Ramanujan sum,
whose value is given as the sum of $d\,\mu(D/d)$ ($\mu=$ M\"obius function) 
over all common divisors~$d$ of $D$ and~$n$. 
Therefore, writing $D$ as~$d\,k$ with $k>0$ we find
\begin{equation}
\ve(n) \= \sum_{d|n} \sum_{k=1}^\infty \frac{\mu(k)}{d\,k^2} \= \frac6{\pi^2}\;\s_{-1}(n)\ ,
\end{equation}
where $\s_i(n):=\sum_{d|n}d^i$ as usual.  This finally gives ($z=x{+}\im y$)
\begin{equation}\label{defC} 
C(z) \= \frac{12}{\pi y}\,\sum_{n=1}^\infty  \bigl(\s_{-1}(n)\+2\pi\,\s_1(n)\,y\bigr)\,\ep^{-2\pi ny}\,\cos(2\pi nx)
\= - \frac{12}{\pi y}\,\log\bigl|\eta(z)\bigr| \,-\, \Re\bigl(E_2(z)\bigr)\ ,
\end{equation}
where $\eta(z)=\ep^{\pi\im z/12}\prod(1-\ep^{2\pi\im nz})^{24}$ and 
$E_2(z)=1-24\sum\s_1(n)\,\ep^{2\pi\im nz}$ denote the Dedekind eta-function 
and the quasimodular Eisenstein series of weight~2 on the full modular group, respectively.  
Note that each of the two terms on the right-hand side of~\eqref{defC} is the sum of a constant~$\pm1$ 
and an exponentially small term of the order of $\ep^{-2\pi y}$, 
but that their sum~$C(z)$ decays exponentially as $y{=}\Im(z)$ grows.  
This explains why in the Tables 4 and~5 the convergence of both $U^{(1)}$ and $U^{(2)}$ 
was much faster for $z_2$ ($y_2{=}1.4$) than for~$z_1$ ($y_1{=}0.7$).

Summarizing, we have given an argument suggesting that the ``correct" refinement of the truncated sum~$U(z,R)$ should be given by
\begin{equation}\label{defUinf} 
U^{(\infty)}(z,R) \= U^{(1)}(z,R) \ +\ \frac{C(z)}R
\end{equation}
with $C(z)$ defined by~\eqref{defC} for $z$ in the complex upper half-plane.  
Table~6 and the graph in Figure~16 both confirm that this improved version 
of the previous functions $U^{(i)}(z,R)$ ($i=0,1,2$) does indeed converge very much faster
 to its limiting value~$U(z)$ than any of them did, with the error now decaying faster than~$1/R$.  
(Both the numerical experiments and a heuristic argument suggest that the true order of magnitude 
of this difference should be something more like $(\log R)^2/R^2$.)
Finally, the graph in Figure~17 gives one last improvement. 
Here, we have replaced the function $U^{(\infty)}(z,R)$ by a function $\overline U(z,R)$
defined as the average value of the numbers $U^{(\infty)}(z,R')$ for $R'$ 
in the (somewhat arbitrarily chosen) interval $[2R/3,R]$.
The calculations up to the same limit $R{=}10^6$ as before 
now yield at least 12 significant digits rather than the original~7. 
\begin{table}[htbp]
{\small
\begin{center}
\begin{tabular}{|c|ccccc|}
\hline
$R$ & $10\,000$ & $50\,000$ & $100\,000$ & $500\,000$ & $1\,000\,000$ \\
\hline
$\rs U^{(\infty)}(z_1,R)\rs\phantom{\Big|}$ & 
52.2450557618857 \rs&\rs 
52.2450552157850 \rs&\rs 
52.2450552237500 \rs&\rs 
52.2450552285072 \rs&\rs 
52.2450552288639 \rs\\
$\rs U^{(\infty)}(z_2,R)\rs$ & 
52.2450554614623 \rs&\rs 
52.2450552195214 \rs&\rs 
52.2450552255882 \rs&\rs 
52.2450552288339 \rs&\rs 
52.2450552288596 \rs\\
\hline
\end{tabular}
\caption{Values of the finally refined truncated sums for the potential, at $z_1=0.1+0.7\,\im$ and $z_2=-1/z_1$}
\end{center}
}
\vspace{-5mm}
\end{table}
\begin{figure}[htbp]
\begin{minipage}{0.31 \paperwidth}
   \includegraphics[scale=0.31]{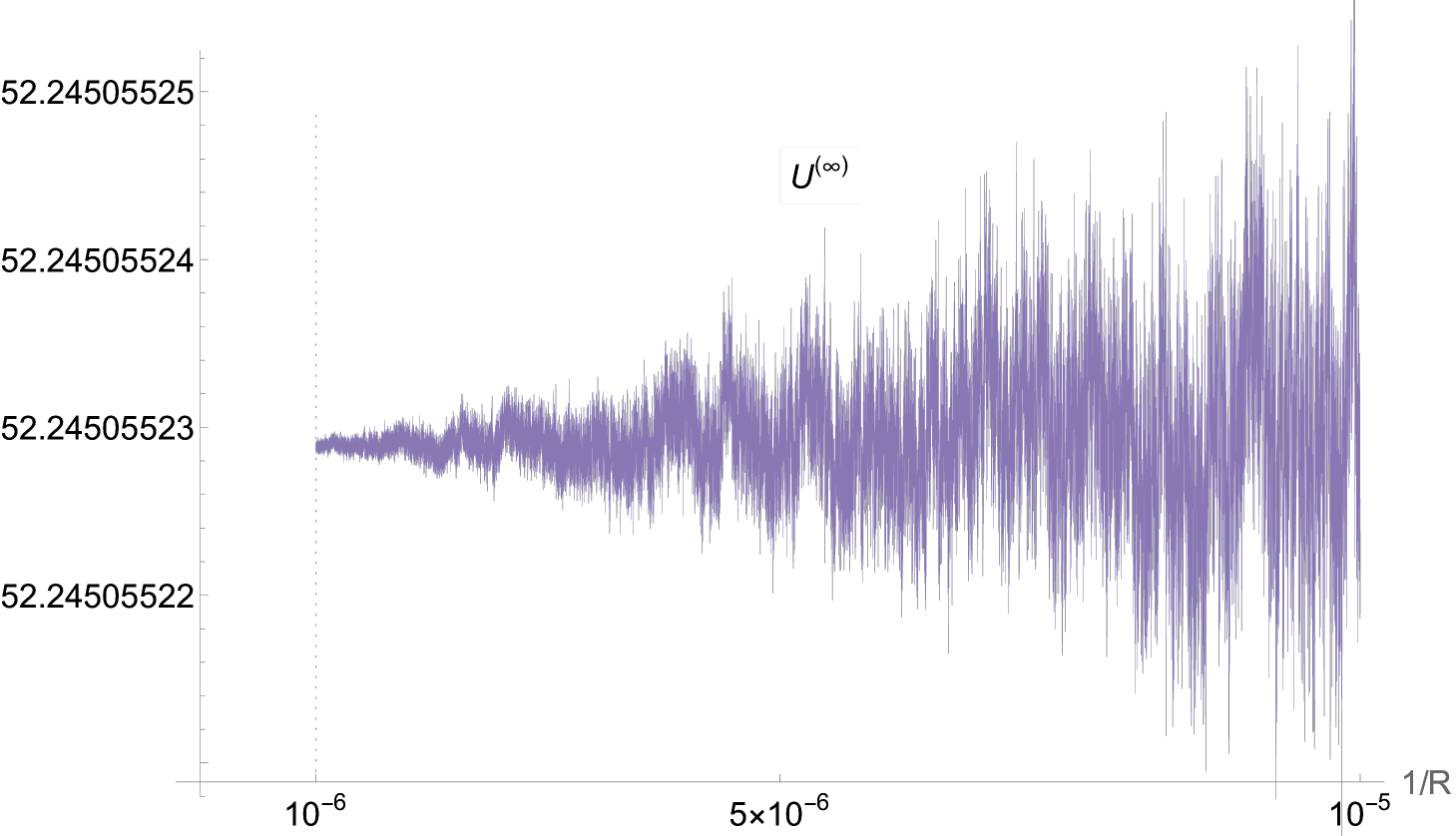} 
   \caption{Plot of the finally refined truncated sum $U^{(\infty)}(z_1,1/R)$}
\end{minipage}
\qquad\qquad\qquad
\begin{minipage}{0.31\paperwidth}
   \includegraphics[scale=0.31]{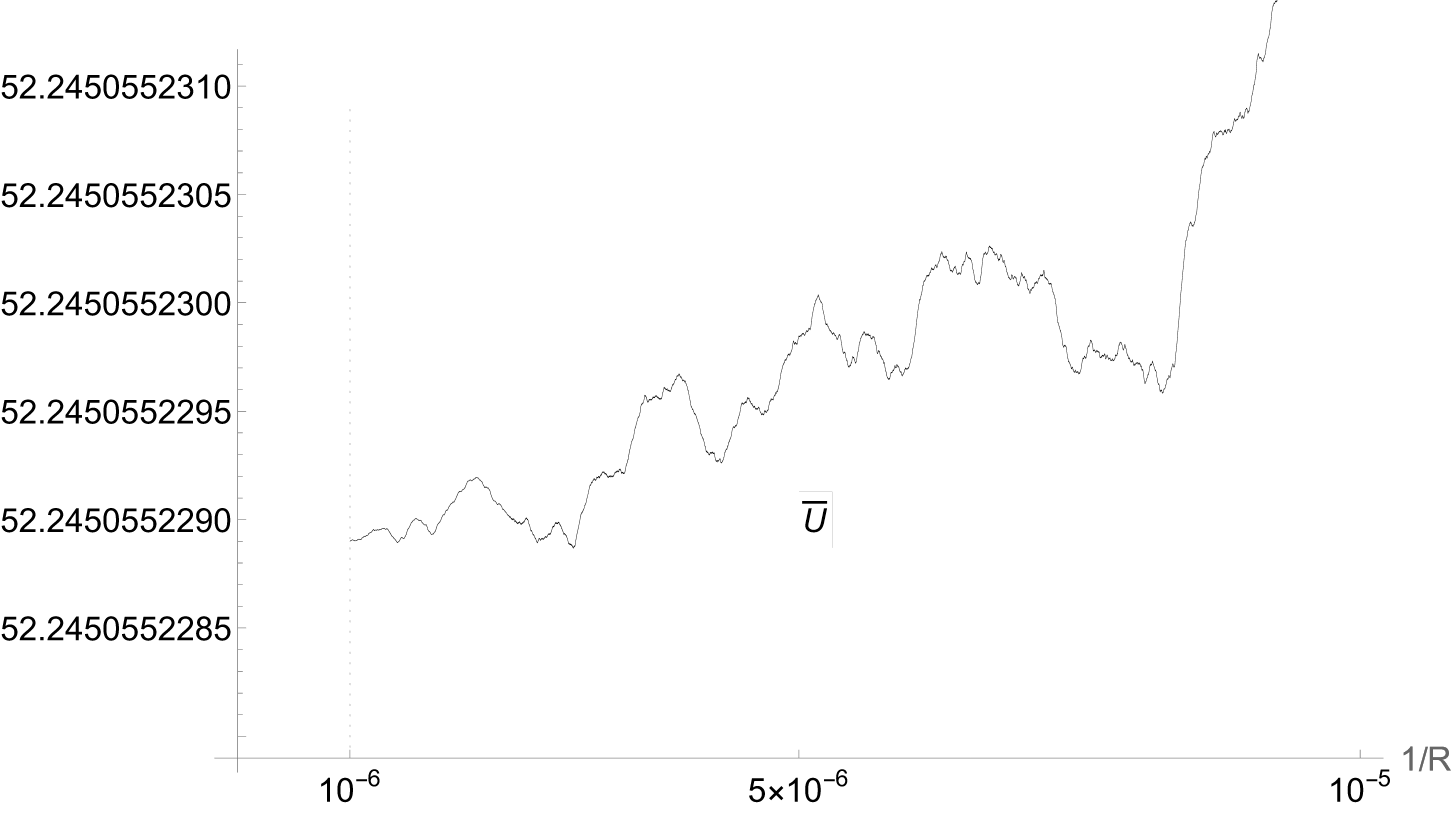}
   \caption{Plot of the averaged truncated sum $\overline{U}(z_1,1/R)$}
\end{minipage}
\end{figure}
\vspace{-1mm}

For the sake of honesty it should perhaps be mentioned that, if our goal were simply to obtain 
a better convergence of $U(z,R)$ to~$U(z)$ as~$R\to\infty$, then we could have avoided 
the whole discussion of the ``right" linear correction term~$C/R$ to~\eqref{defU0}.
One can simply use a least-squares fit to obtain~$C$ numerically from our tabulated values 
(as we in fact did originally, with results that were not all that much worse than~$U^{(\infty)}$).
Alternatively, at the cost of a little loss of accuracy, one can replace $U^{(0)}(R)$ by the expression 
$2\,U^{(0)}(R)-U^{(0)}(R/2)$, which eliminates any linear term in~$1/R$ and is unchanged 
by employing $U^{(1)}$, $U^{(2)}$, or $U^{(\infty)}$ instead of $U^{(0)}$.  
However, our analysis leading to the final correction term as given by
\eqref{defU0}, \eqref{defU1} and \eqref{defUinf} with~\eqref{defC}
is mathematically interesting and seemed worth giving, 
especially in view of the unexpected occurrence
of the nearly modular functions $\log\bigl|\eta(z)\bigr|$ and $\Re\bigl(E_2(z)\bigr)$.
\\[-20pt]

\subsection*{Acknowledgments}
\vspace{-4mm}
\noindent
I am grateful to Don Zagier for illuminating discussions and
to Hermann Karcher for helpful comments.
I also thank Luca Romano for collaboration at an early stage of this project
and Axel Kleinschmidt for pointing out some literature.
%

%

\end{document}